\documentclass[prd,aps,amsmath,amssymb, preprintnumbers, reprint,longbibliography,superscriptaddress,nofootinbib]{revtex4-1}

\usepackage{bm,graphicx,wasysym}
\usepackage{xcolor}
\usepackage{color}
\usepackage{lipsum}

\allowdisplaybreaks

\newcommand\sumint[1]{\int\kern-1.5em\sum\nolimits_{#1}}

\def\({\left(}
\def\){\right)}
\def\[{\left[}
\def\]{\right]}
\def\d{\mathrm{d}}

\newcommand{\f}[2]{\frac{#1}{#2}}
\def \bal#1\eal  {\begin{align} #1 \end{align}}
\newcommand{\be} {\begin{equation}}
\newcommand{\ee} {\end{equation}}
\newcommand{\bc}{\begin{center}}
\newcommand{\ec}{\end{center}}
\newcommand{\bim} {\begin{itemize}[noitemsep]}

\newcommand{\eim} {\end{itemize}}

\newcommand{\nn} {\nonumber\\}
\newcommand{\eref}[1]{Eq.~(\ref{#1})}

\newcommand{\si}{{\sigma}}
\newcommand{\li}{{\lambda}}

\newcommand{\oi}{\omega}

\begin{document}

\hfill {{\footnotesize USTC-ICTS/PCFT-23-16}}

\title{Boson Star Superradiance}

\author{He-Yu Gao}
\email{gao\_he\_yu@mail.ustc.edu.cn}
\affiliation{School of Physical Sciences, University of
Science and Technology of China, Hefei, Anhui 230026, China}

\author{Paul M.~Saffin}
\email{paul.saffin@nottingham.ac.uk}
\affiliation{School of Physics and Astronomy, University of Nottingham,\\
University Park, Nottingham NG7 2RD, United Kingdom}

\author{Yi-Jie Wang}
\email{yjwang@mail.ustc.edu.cn}
\affiliation{School of Physical Sciences, University of
Science and Technology of China, Hefei, Anhui 230026, China}

\author{Qi-Xin Xie}
\email{xqx2018@mail.ustc.edu.cn}
\affiliation{Interdisciplinary Center for Theoretical Study, University of
Science and Technology of China, Hefei, Anhui 230026, China}

\author{Shuang-Yong Zhou}
\email{zhoushy@ustc.edu.cn}
\affiliation{Interdisciplinary Center for Theoretical Study, University of
Science and Technology of China, Hefei, Anhui 230026, China}
\affiliation{Peng Huanwu Center for Fundamental Theory, Hefei, Anhui 230026, China}

\begin{abstract}

Recently, it has been realized that in some systems internal space rotation can induce energy amplification for scattering waves, similar to rotation in real space. Particularly, it has been shown that energy extraction is possible for a Q-ball, a stationary non-topological soliton that is coherently rotating in its field space. In this paper, we generalize the analysis to the case of boson stars, and show that the same energy extraction mechanism still works for boson stars.  

\end{abstract}

\maketitle

\section{Introduction}

Boson stars are localized stationary solutions composed of scalar and gravitational fields \cite{Kaup:1968zz}. They are stable thanks to the balance between the dispersion of the scalar modes and the gravitational attraction. Depending on the mass and couplings of the scalar field, boson stars can attain macroscopic sizes and possess large energies, and even exhibit high gravitational compactness. Many varieties of boson stars have been uncovered over the years, encompassing diverse aspects such as scalar self-interactions, scalar-gravity couplings, novel constituents and dynamics, among others \cite{Ruffini:1969qy,Colpi:1986ye,Friedberg:1986tq,Jetzer:1989av,Henriques:1989ar,Seidel:1991zh,Silveira:1995dh,Schunck:1996he,Yoshida:1997qf,Schunck:1999zu,Bernal:2009zy, Herdeiro:2019mbz}; see \cite{Liebling:2012fv,Visinelli:2021uve} for a review. Furthermore, boson stars can serve as dark matter sources \cite{Lee:1995af,Chavanis:2011zi,Rindler-Daller:2011afd,Eby:2015hsq,March-Russell:2022zll,Navarro-Boullosa:2023bya} and black hole mimickers \cite{Torres:2000dw,Guzman:2009zz,Olivares:2018abq, Herdeiro:2021lwl}. They and their binary systems can emit distinctive gravitational radiation potentially detectable by future gravitational wave experiments \cite{Berti:2006qt,Palenzuela:2007dm,Cardoso:2016oxy,Sennett:2017etc,Palenzuela:2017kcg,Dietrich:2018bvi,Bezares:2018qwa,Croon:2018ybs,Choi:2019mva,Bezares:2022obu,Jaramillo:2022zwg,Croft:2022bxq}. 

The term superradiance was originally coined by Dicke \cite{Dicke:1954zz} in the study of emission enhancement of radiation in a coherent medium. Subsequently, it has been widely employed to describe a diverse array of phenomena characterized by enhanced radiation. One of the prominent examples is (real space) rotational superradiance, which refers to the ability of a rotating body to transfer its energy and angular momentum to the surrounding environment via its scattering of the waves impinging on it \cite{Zeld1, Zeld2}. As Kerr black holes are just such rotating bodies, black hole superradiance has been widely studied in the field of gravitational physics and astrophysics over the past decades \cite{Misner:1972kx, Cardoso:2004hs, Dolan:2007mj, Hartnoll:2008vx, Casals:2008pq, Arvanitaki:2009fg, Arvanitaki:2010sy, Pani:2012vp, Witek:2012tr, Herdeiro:2013pia, Yoshino:2013ofa, Basak:2002aw, Richartz:2014lda, Torres:2016iee, Benone:2015bst, Konoplya:2016hmd, Hod:2016iri, Baryakhtar:2017ngi, East:2017ovw, Rosa:2017ury, Cardoso:2018tly, Degollado:2018ypf, Zhang:2020sjh, Mehta:2021pwf}. Superradiance in rotating stars has recently been discussed in \cite{Richartz:2013unq, Cardoso:2015zqa, Cardoso:2017kgn, Day:2019bbh, Chadha-Day:2022inf}. Reviews about superradiance can be found in \cite{Bekenstein:1998nt, Brito:2015oca}. 

Similar to boson stars, Q-balls \cite{Rosen:1968mfz,Friedberg:1976me,Coleman:1985ki,Lee:1991ax} are localized stationary solutions of scalar fields in flat spacetime. These intriguing structures may arise naturally during the early stages of the universe \cite{Enqvist:1997si,Kasuya:1999wu,Multamaki:2002hv,Harigaya:2014tla, Zhou:2015yfa}. Furthermore, Q-balls have also been postulated as plausible candidates for dark matter \cite{Kusenko:1997si,Banerjee:2000mb,Roszkowski:2006kw,Shoemaker:2009kg,Kasuya:2012mh}, and can have composite nonlinear structures \cite{Copeland:2014qra,Xie:2021glp,Hou:2022jcd}. Recently, Ref \cite{Saffin:2022tub} has discovered a novel mechanism to extract energy from a Q-ball by incident waves, which was referred to as Q-ball superradiance. The energy extraction is possible owing to the Q-ball's coherent internal rotation in field space and the coupling of the two inherent scalar modes. While Ref \cite{Saffin:2022tub} defined the energy amplification with the total energy in a far-away region and identified the amplification criteria accordingly, Ref \cite{cardosoSub} pointed out that this does not coincide with the energy amplification defined using the flux to and from infinity, due to the difference between the group velocities of the two coupled modes \cite{cardosoSub}. Indeed, with the flux definition, a Zel'dovich-like amplification criterion can be found when there is only one ingoing mode. The difference between the two definitions presents an intriguing feature of such a system. Ref \cite{cardosoSub} has also argued that generic time-periodic solitons such as Newtonian boson stars are prone to energy extraction. In view of the resemblance between Q-balls and (relativistic) boson stars, it is reasonable to anticipate a similar mechanism applies to boson stars as well. The purpose of this paper is to confirm this anticipation and explicitly demonstrate that the internal rotation of a (relativistic) boson star also allows the same kind of energy extraction. 

The paper is organized as follows. Section \ref{sec:model} presents the model we will focus on, setting up the stage. In Section \ref{sec:bosonstar}, we compute the background boson star solutions by formulating it in a manner suitable for solving using a 1D shooting method. In Section \ref{sec:BSsuper}, we investigate the superradiant amplification of scattering waves impinging on a boson star and delineate the criteria for the energy extraction to occur. We summarize in Section \ref{sec:summary}. 

\vspace{-10pt}
\section{Model}
\label{sec:model}

As mentioned above, for our novel energy extraction mechanism to work, we need at least two interacting degrees of freedom. So we consider a complex scalar field $\Phi$ minimally coupled to the Einstein-Hilbert action:
\be
{S}=\int \mathrm{d}^4 x \sqrt{-g}\left[\frac{{R}}{16 \pi G}-\nabla^\mu \Phi^\dagger \nabla_\mu \Phi-V\left(|\Phi|^2\right)\right] ,
\ee
where the potential $V$ is to be specified later. Depending on convenience, we shall also use the reduced Planck mass $M_P=(8\pi G)^{-1/2}$. The equations of motion for this system are given by
\bal
R_{\mu \nu}-\frac{1}{2} g_{\mu \nu} R&=8 \pi G T_{\mu \nu},\\
\nabla^\mu \nabla_\mu \Phi &=\frac{\mathrm{d} V\left(|\Phi|^2\right)}{\mathrm{d} | \Phi|^2} \Phi,
\eal
where $|\Phi|^2= \Phi^\dagger \Phi$ and the energy-momentum tensor of the bosonic field $\Phi$ is given by
\be
T_{\mu \nu}=2\nabla_{(\mu} \Phi^\dagger \nabla_{\nu)} \Phi - g_{\mu \nu}\left(\nabla^\gamma \Phi^\dagger \nabla_\gamma \Phi+V\right).
\ee
Here the symmetrization is defined as $B_{(\mu\nu)}=(B_{\mu\nu}+B_{\nu\mu})/2!$. The action is invariant under a global $U(1)$ symmetry $\Phi \rightarrow \Phi e^{i \kappa}$, $\kappa$ being a real constant, which leads to a conserved Noether current $\nabla_\mu J^\mu=0$ with $J_\mu= i\left(\Phi^\dagger \nabla_\mu \Phi-\nabla_\mu \Phi^\dagger \Phi\right)$.

\section{Boson star}
\label{sec:bosonstar}

The energy extraction mechanism results from rotation in the internal space, so it suffices to demonstrate it in spherical symmetry in real space.  To this end, we adopt the following ansatz for the scalar and the metric: 
\bal
\label{ansatz01}
\Phi_B(r, t)&=\phi_B(r) e^{-i \omega_B t},
\\
\label{ansatz02}
\mathrm{d} s^2&=-e^{v(r)} \mathrm{d} t^2+e^{u(r)} \mathrm{d} r^2+r^2\d\Omega^2,
\eal
where $\d \Omega^2 = \mathrm{d} \vartheta^2+\sin ^2 \vartheta \mathrm{d} \varphi^2$. $\phi_B(r)$, $v(r)$ and $u(r)$ are three real radial functions to be solved with the field equations. With this ansatz, the field equations read
\bal
\label{eurho1}
e^{-u}\left(\frac{u^{\prime}}{r}-\frac{1}{r^2}\right)+\frac{1}{r^2} & =8 \pi G \rho_\Phi, \\
e^{-u}\left(\frac{v^{\prime}}{r}+\frac{1}{r^2}\right)-\frac{1}{r^2} & =8 \pi G p_{\Phi}, \\
\phi_B^{\prime \prime}+\left(\frac{2}{r}+\frac{v^{\prime}-u^{\prime}}{2}\right) \phi_B^{\prime} & =e^u\left( V_{\phi_B^2}-e^{-v} \omega_B^2\right) \phi_B,
\label{eurho3}
\eal
where $V_{\phi_B^2}=\d V(\phi_B^2)/\d \phi_B^2$, a prime ${}'$ denotes a derivative with respect to $r$, and the energy density $\rho_\Phi$ and radial pressure $p_{\Phi}$ are given respectively by
\bal
\rho_\Phi & =e^{-v} \omega_B^2 \phi_B^2+e^{-u}\left(\phi_B^{\prime}\right)^2+V , \\
p_{\Phi} & =e^{-v} \omega_B^2 \phi_B^2+e^{-u}\left(\phi_B^{\prime}\right)^2-V .
\eal
Defining a radius-dependent mass quantity $\mathcal{M}(r)$ via $e^{-u}=1-2G \mathcal{M}(r)/ r$, it is easy to see \eref{eurho1} can be re-written as 
\be
\frac{\mathrm{d} \mathcal{M}(r)}{\mathrm{d} r}=4 \pi r^2 \rho_\Phi .
\ee
So one may loosely refer to $\mathcal{M}(R)=\int_0^R 4 \pi r^2 \d r \rho_\Phi $ as the energy of the boson star within radius $R$. From this definition, we see that the total energy of the boson star is simply the ADM mass of the asymptotically flat spacetime 
\be
M=\lim_{r \rightarrow +\infty} \mathcal{M}(r) .
\ee

Now let us parametrize the scalar potential to be used in this paper. We assume that the potential can be expanded perturbatively around $\Phi=0$. Due to the U(1) symmetry, the leading terms of the effective potential are as follows:
\be
\label{potential}
V(|\Phi|^2)=\mu^2|\Phi|^2+\frac{\lambda}{2}|\Phi|^4+\frac{\nu}{3}|\Phi|^6 +\cdots.
\ee
A preliminary numerical survey seems to suggest that significant energy amplification is possible if $\lambda<0$, a strong resemblance to the case of Q-balls. Although its effects on the energy extraction is negligible, a small $|\Phi|^6$ term can prevent the vacuum from decaying quantum mechanically. We shall neglect the higher order terms in the following explicit computations.

For a given $\omega_B$, Eqs.~(\ref{eurho1}-\ref{eurho3}) with appropriate boundary conditions can be numerically solved by viewing it as an ``initial" value problem with $r$ playing the role of ``time''. That is, one can start at a small $r$ with appropriate ``initial'' conditions and integrate to a large $r$ to satisfy certain ``final'' conditions. However, a two-dimensional shooting method is needed to directly solve this system, corresponding to choosing appropriate values of $\phi_B$ and $v$ at $r=0$. Fortunately, the time scaling symmetry of the solutions allows us to cast Eqs.~(\ref{eurho1}-\ref{eurho3}) into a system that can be solved by a one-dimensional shooting method, which is much easier to implement. That is, if a solution is given by Eqs.~(\ref{ansatz01}-\ref{ansatz02}), simultaneously scaling $t\to \li t$ and $e^v\to \li^2e^v$ gives another solution. This allows us to eliminate $\omega_B$ from the equations of motion. 

\begin{figure}
    \centering
    \includegraphics[width=8cm]{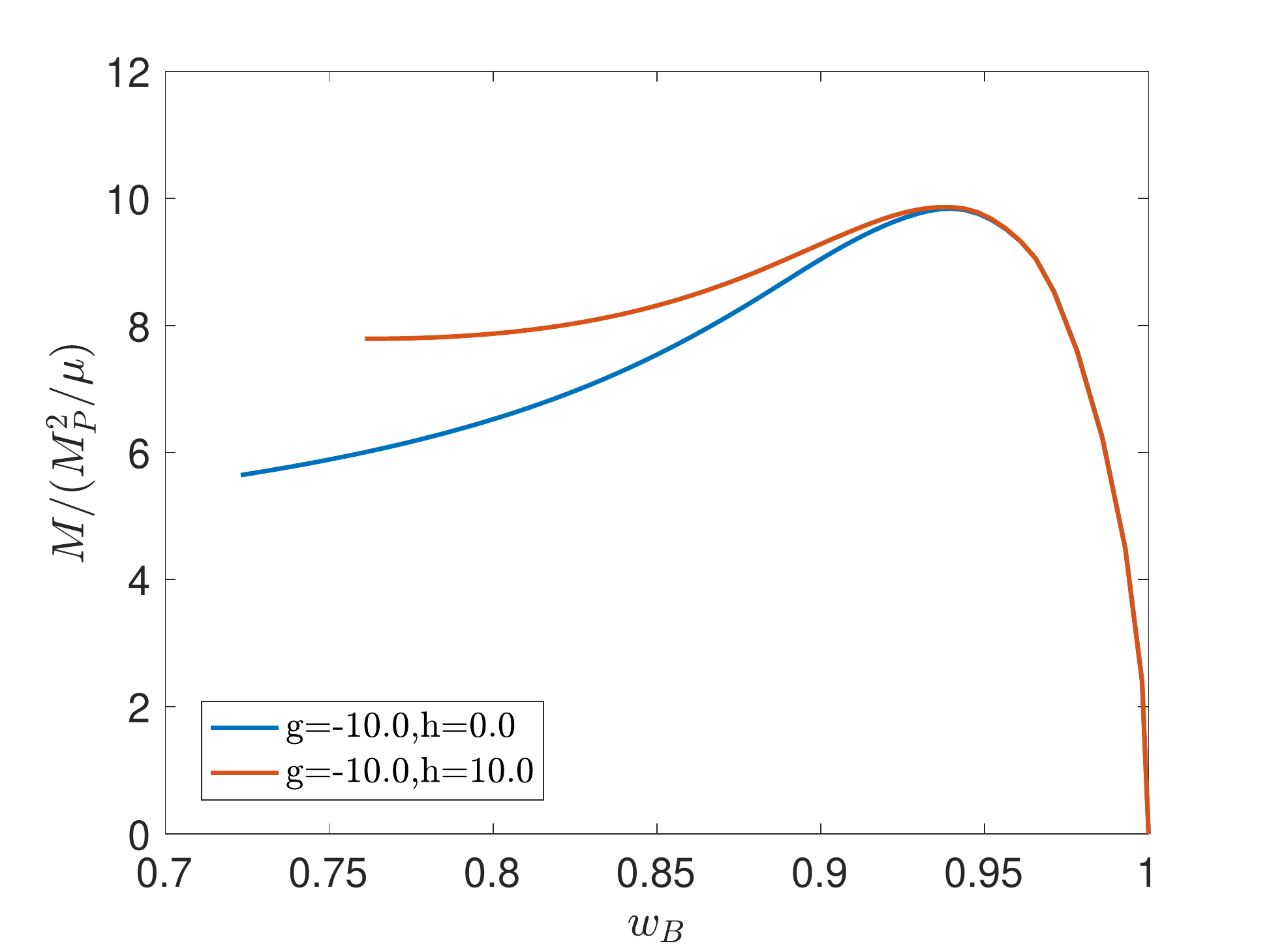}
    \caption{Relation between the boson star mass $M$ and internal rotational frequancy $w_B$. The potential parameters $g$ and $h$ are defined in \eref{dimlessdef}.}
    \label{fig:m2omega}
\end{figure}

To this end, we shall replace the $v$ variable with the dimensionless variable
\be
\sigma=e^{\frac{u+v}{2}}/w_B, 
\ee
where 
\be
w_B ={\omega_B}/{\mu} .
\ee
Together with the dimensionless variables
\be
\label{dimlessdef}
x=r\mu,~~ f=\f{\phi_B}{M_P}, ~~
g=\frac{\lambda M_P^2}{ \mu^2}, ~~h=\frac{\nu M_P^4}{\mu^2},
\ee
 the equations of motion (\ref{eurho1}-\ref{eurho3}) can be written as
\bal
 \dot{u} &=\frac{1-e^u}{x}+x e^u  f^2\left(\frac{e^u}{\sigma^2}+F\right)+x \dot{f}^2 ,
 \\ 
  \dot{\sigma} &=x\left(  \frac{e^{2 u}  f^2}{\sigma}+\sigma \dot{f}^2\right),
  \\ 
  \ddot{f} &=-\left(1+e^u-x^2 e^u  f^2 F\right) \frac{\dot{f} }{x} +e^u\left( F_1-\frac{e^u}{\sigma^2}\right)  f ,
\eal
where a dot $\dot{}$ denotes a derivative with respect to $x$ and we have defined
\be
F=1+\frac{1}{2} g  f^2+\frac{1}{3}h f^4,~~ F_1=1+g  f^2+h f^4.
\ee
The appropriate boundary conditions for an asymptotically flat boson star solution are
\bal
 u(0)&=0, ~~ u(+\infty)=0,~~  \si(0)=\si_0 ,~~ \si(+\infty)=1/w_B ,
 \nn
 f(0)&= f_0 ,~~  \dot{f}(0)=0 ,~~  f(+\infty)=0 .
\eal
In this formulation, we only need to shoot to find the correct value of $f_0$. The boundary value of $\si(+\infty)$ does not require a shooting procedure, as it merely gives the internal rotational frequency of the complex scalar of the boson star: $\si(+\infty)\to 1/w_B$. In Figure \ref{fig:m2omega}, we have computed the relation between the boson star mass $M$ and the internal space rotational frequency of the boson star $w_B$ for two example sets of potential parameters.

\section{Scattering off a boson star}

\label{sec:BSsuper}

Having found the boson star solution, we now proceed to scatter waves on to the boson star and examine how the energy of the outgoing waves
changes, compared to the ingoing waves. To this end, we view the scattering waves as a small perturbation $\Theta(t, r)$ around the boson star background $\Phi_B(t, r)=\phi_B(r)e^{-i\omega_Bt}$:
\be
\Phi(t, r)=\Phi_B(t, r)+ \Theta(t, r).
\ee
To leading order, the equation of motion for $\Theta$ is given by
\be
\square \Theta  -\left.\frac{\partial^2 V}{\partial \Phi^{\dagger} \partial \Phi}\right|_{\Phi_B} \Theta -\left.\frac{\partial^2 V}{\partial \Phi^{\dagger} \Phi^{\dagger}}\right|_{\Phi_B} \Theta^{\dagger} =0,
\ee
where $\square=e^{-u} (\partial_{rr}+\frac{v'-u'}{2}\partial_r+\frac{2}{r}\partial_r )-e^{-v}\partial_{tt}$.
We shall neglect the back-reaction of the scattering waves on the background geometry. Due to the $\Theta^{\dagger}$ term, the two real modes of the perturbative complex scalar are coupled, which, as we will see, is fundamental for our energy extraction mechanism to work. For the potential (\ref{potential}), we have
\be
\label{ThiEq}
\square \Theta -U(r)\Theta -W(r)e^{-2i\omega_B t}\Theta^\dagger=0,
\ee
where we have defined
\bal
U(r)&=\mu^2+2\lambda|\phi_B|^2+3\nu|\phi_B|^4,
\\
W(r)
&=(\lambda+2\nu|\phi_B|^2)|\phi_B|^2.
\eal
So we see that the coherent internal rotation of the boson star provides a driving factor $e^{-2i\omega_B t}$ in \eref{ThiEq}, another important ingredient for the energy amplification mechanism to work.

Let us first consider the simplest case of a quadratic potential,
\be
V\left(|\Phi|^2\right)=\mu^2|\Phi|^2.
\ee
The quadratic potential can only support mini-boson stars, whose masses are limited by the Kaup bound $M_{\rm Kaup} \approx 15.9 M_{P}^2/\mu$ \cite{Kaup:1968zz}. Also, mini-boson stars can only achieve relatively low compactness. Much more massive and compact boson stars can be obtained with interacting potentials. 
The free quadratic potential can be obtained by setting $g=h=0$ in \eref{ThiEq}, which leads to $W=0$ and $\Theta$ satisfying
\be
\square \Theta-\mu^2\Theta=0.
\ee
In this case, there is no mode mixing, so the new energy amplification mechanism is not at work here, which we have confirmed with explicit numerical computations.

\begin{figure}
    \centering
    \includegraphics[width=3.45in]{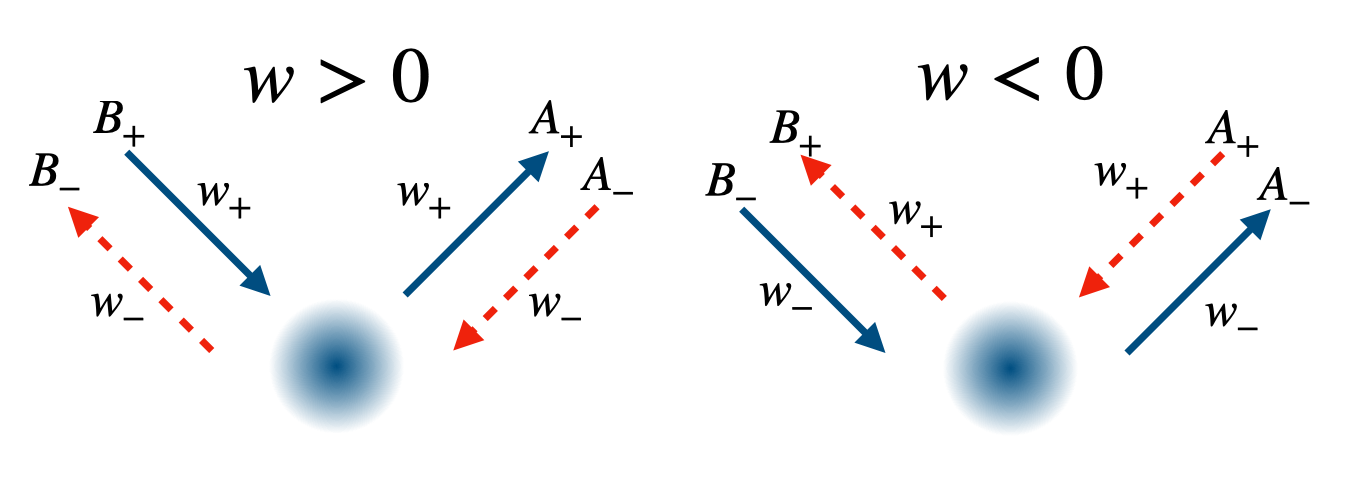}
    \caption{Waves scattering on and off a boson star (cf.~\eref{asympform}). The frequencies of the two modes are $w_\pm=w_B\pm w$. We choose $w_B>0$ (thus a positively charged boson star) and use solid blue (dashed red) lines to represent positive (negative) charges. }
    \label{fig:ABwaves}
\end{figure}

For the generic case of an interacting scalar field theory, which supports more massive boson stars and thus is more interesting phenomenologically, scattering waves can extract energy from a boson star. To see this, let us
consider the minimal case of two scattering modes:
\be
\Theta=\eta_{+}(\omega, r) e^{-i \omega_{+} t}+\eta_{-}(\omega, r) e^{-i \omega_{-} t} ,
\ee
where 
\be
\omega_{ \pm}=\omega_B \pm \omega .
\ee
The equations of motion for $\eta_\pm$ are
\bal
\label{mainpertEq0}
& e^{-u}\left[\eta_\pm''+\(\frac{v'-u'}{2}+\frac{2}{r}\)\eta_\pm'\right] 
\nn
&~~~~~~~~~~~~~ +\(e^{-v}\omega_\pm^2-U(r)\)\eta_\pm-W(r)\eta_\mp^\dagger=0.
\eal
Defining 
\be
\chi_\pm= \f{\eta_\pm}{M_P},~~ w_\pm = \f{\omega_\pm}{\mu},~~q=v-u
\ee
and using the dimensionless variables defined around \eref{dimlessdef}, \eref{mainpertEq0} becomes
\be
\label{mainEq1}
\ddot{\chi}_\pm+\frac{x\dot{q}+4}{2x}  \dot{\chi}_\pm+(e^{-q} w_\pm^2 -U e^u) \chi_\pm-W e^u \chi_\mp^{\dagger}=0,
\ee
where again a dot is a derivative with respect to $x$ and we have defined
\be
U(x)=1+2g f ^2+3h f ^4, ~~W(x)=g f ^2+2h f ^4.
\ee
Asymptotically at large $x$, the scattering waves take the form
\be
\label{asympform}
\chi_\pm(w,x\to+\infty) \to\frac{ A_\pm}{k_\pm x}e^{ik_\pm x}+\frac{B_\pm}{k_\pm x}e^{-ik_\pm x},
\ee
where we have defined $k_\pm=(w_\pm^2-1)^{1/2}$. This means that for propagating waves, we need $|w_\pm|=|w_B\pm w|>1$, with the 1 coming from the scalar mass $\mu^2$ in dimensionful variables. Without loss of generality, we shall assume $w_B>0$, and thus we have a mass gap for the frequencies
\be
|w|>w_B+1.
\ee
As we have chosen $k_\pm\geq 0$, whether $A_\pm$ and $B_\pm$ represent incoming or outgoing waves depends on the sign of $w$; see Figure \ref{fig:ABwaves} for a quick reference.

Before numerically solving \eref{mainEq1} for scattering waves with the asymptotical form (\ref{asympform}), we would like to point out an important property of this kind of scattering, which is that the ``particle number'' is conserved in the process. 
To see this, we define
\be
Y=xe^{q/4}\chi_+,~~Z=xe^{q/4}\chi_-^\dagger
\ee
and
\be
\varrho_\pm=-(\frac{\ddot{q}}{4}+\frac{\dot{q}^2}{16}+\frac{\dot{q}}{2x})+e^{-q} w_\pm^2-U e^u,~~~
\varpi=-We^u,
\ee
and then \eref{mainEq1} and its complex conjugate can be written as
\bal
\label{Yequ}
Y''+\varrho_+(x)Y+\varpi(x)Z&=0,
\\
\label{Zequ}
Z''+\varrho_-(x)Z+\varpi(x)Y&=0.
\eal
It is easy to see that the equations of motion are invariant under a global U(1) transformation  
$$
\left(\begin{array}{l}
Y \\
Z
\end{array}\right) \rightarrow\left(\begin{array}{l}
Y \\
Z
\end{array}\right) e^{i \alpha},
$$
where $\alpha$ is a real constant. This U(1) symmetry on the perturbative field is inherited from the U(1) symmetry of the full field. The Noether charge for this symmetry is given by 
\bal
\mathcal{N} &=i\left(Y^{\dagger} \partial_x Y-Y \partial_x Y^{\dagger}\right)+i\left(Z^{\dagger} \partial_x Z-Z \partial_x Z^{\dagger}\right)
\eal
and we have $\d \mathcal{N}/\d x=0$. At $x=0$, it is easy to see that $\mathcal{N}=0$. Asymptotically, as $x\to +\infty$, we have
\be
\mathcal{N} \to -2\left(\frac{|A_+|^2}{k_+}-\frac{|B_+|^2}{k_+}+\frac{|B_-|^2}{k_-}-\frac{|A_-|^2}{k_-}\right)e^{{q}/{2}},
\ee
so conservation of $\mathcal{N}$ implies that in the scattering the ``particle number'' of the ingoing waves is the same as that of the outgoing waves:
\be
\label{particleCons}
\frac{|A_+|^2}{k_+}+\frac{|B_-|^2}{k_-}=\frac{|A_-|^2}{k_-}+\frac{|B_+|^2}{k_+},
\ee
which we have confirmed numerically. The reason why the above equation corresponds to the particle number conservation in the scattering can be understood as follows. The radial flux of the Noether current at infinity over a unit area and averaged over time is $J_r= \f2{k_+r^2}\(|B_+|^2-|A_+|^2\)+\f2{k_-r^2}\(|B_-|^2-|A_-|^2\)$. By identifying a unit charge as a particle and recalling the meanings of these coefficients as in Figure \ref{fig:ABwaves}, we find that the number of the outgoing particles is equal to that of the ingoing ones. (Recall that the flux of a spherical wave of the form $Ce^{ikr-i\oi t}$ is $-2k|C|^2$.)

\begin{figure}
\centering
    \includegraphics[width=0.45\textwidth]{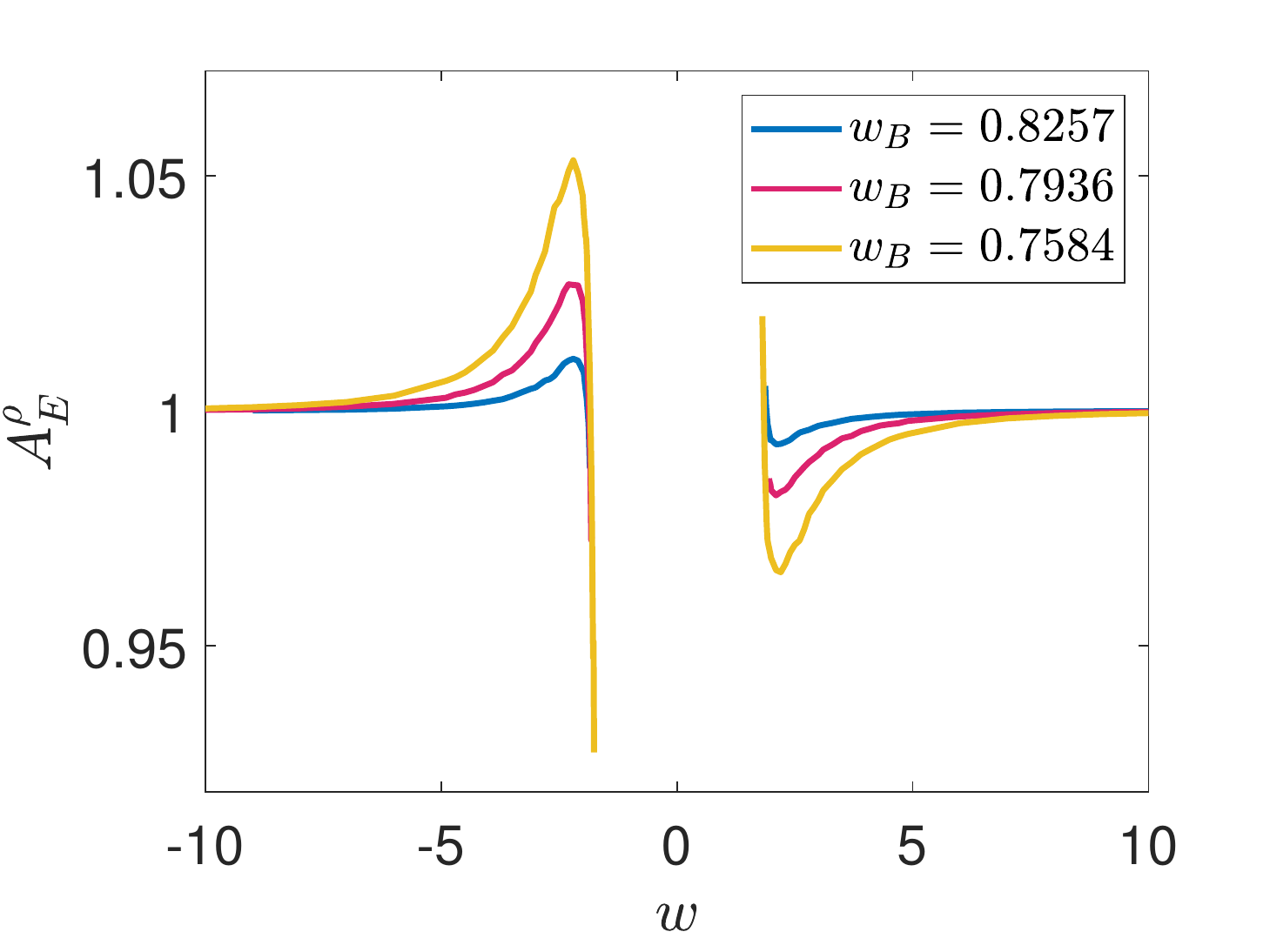}
    \caption{Energy amplification factor $\mathcal{A}^{\rho}_E$ (defined with the total scalar energy) with only the $+$ ingoing modes. $w_B$ is the frequency of the boson star internal rotation. The $+$ and $-$ scattering waves have frequency $w_+=w_B+w$ and $w_-=w_B-w$ respectively. The gap around $w=0$ is due to the fact that the scalar has a mass. The potential parameters are $g=-10$ and $h=0.5$.}
    \label{fig:rho1D}
\end{figure}

We also want to define the energy amplification factor for the asymptotic waves. Since the system contains two modes with different group velocity, there are two ways to define the energy amplification. One is to take into account all the scalar energy in an asymptotic spherical shell with interior radius $r_1$ and exterior radius $r_2$
\bal
E_{\circledcirc} & =\frac{1}{r_2-r_1}\int_{r_1}^{r_2}\!\! 4\pi r^2 \mathrm{d} r \, \left< \rho_\Phi \right>
\\
& \propto \frac{w_{+}^2}{k_{+}^2}\left(\left|A_{+}\right|^2+\left|B_{+}\right|^2\right)+ \frac{w_{-}^2}{k_{-}^2}\left(\left|A_{-}\right|^2+\left|B_{-}\right|^2\right),\nonumber
\eal
where $r_2-r_1$ includes at least a full spatial oscillation of the longest wave, $\left<~\right>$ denotes time average over a few oscillations and we have kept the leading order in $1/r$. Using this total energy measure, the energy amplification factor can be defined as the ratio of energy in the outgoing modes against the ingoing modes,
\be
\label{ArhoE}
\mathcal{A}^{\rm \rho}_E=\left|\frac{\frac{w_+^2}{k_+^2}|A_+|^2+\frac{w_-^2}{k_-^2}|B_-|^2}{\frac{w_+^2}{k_+^2}|B_+|^2+\frac{w_-^2}{k_-^2}|A_-|^2}\right|^{{\rm sign}(w)}.
\ee
For this definition of amplification and the ``conservation'' law (\ref{particleCons}), one can immediately see that the criterion that delineates the amplification and attenuation of the total average energy is
\be
\frac{w_+^2}{k_+} = \frac{w_-^2}{k_-} ,
\ee
as this is the point when $\mathcal{A}^{\rm \rho}_E=1$. Another measure one may use, which was first pointed out in \cite{cardosoSub}, is to compute the amplification factor for the energy fluxes to and from spatial infinity:
\bal
{\cal F}_\infty & =\lim_{r\to \infty} 4\pi r^2  \left< T_{tr} \right>
\\
& \propto \frac{w_{+}}{k_{+}}\left|A_{+}\right|^2-\frac{w_{+}}{k_{+}}\left|B_{+}\right|^2+\frac{w_{-}}{k_{-}}\left|A_{-}\right|^2-\frac{w_{-}}{k_{-}}\left|B_{-}\right|^2 \nonumber.
\eal
With this energy flux, we can define another energy amplification factor
\be
\mathcal{A}^{\cal F}_E=\left|\frac{\frac{w_+}{k_+}|A_+|^2-\frac{w_-}{k_-}|B_-|^2}{\frac{w_+}{k_+}|B_+|^2-\frac{w_-}{k_-}|A_-|^2}\right|^{{\rm sign}(w)}.
\ee
For this flux definition, if there is only one mode ingoing (but two modes outgoing), one can find a clear Zel'dovich-like amplification criterion \cite{cardosoSub}. That is, if there is only the $+$ mode ingoing, $\mathcal{A}^{\cal F}_E$ will be greater than 1 if $w_+=w_B+w<w_B$, {\it i.e.}, $w<0$; if there is only the $-$ mode ingoing, $\mathcal{A}^{\cal F}_E$ will be greater than 1 if $w_-=w_B-w<w_B$, {\it i.e.}, $w>0$.

\begin{figure}
\centering
    \includegraphics[width=0.45\textwidth]{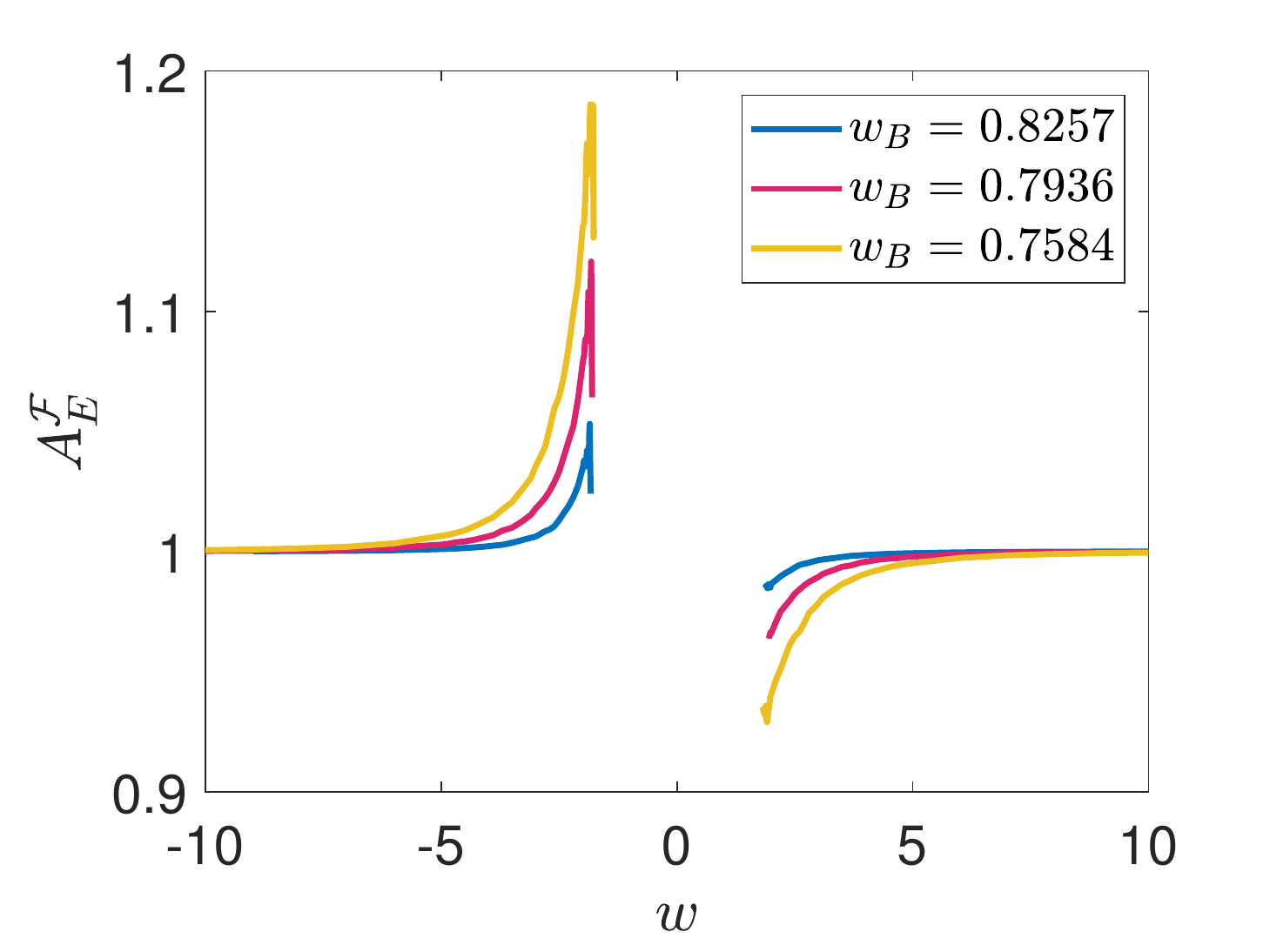}
    \caption{Same as Figure \ref{fig:rho1D} but with the energy amplification factor $\mathcal{A}^{\cal F}_E$ (defined with the flux scalar energy).}
    \label{fig:F1D}
\end{figure}

For the case of a single mode (both ingoing and outgoing), the two definitions of amplification would coincide. However, for the generic case where two modes are involved, they approximately agree with each other far away from the mass gap, but differ near the gap ({\it i.e.}, when $|w_\pm|=(k_\pm^2+1)^{1/2}$ is close to 1). As we will see, if there is only one mode ingoing (but two modes outgoing), $\mathcal{A}^{\rho}_E$ is typically greater than $\mathcal{A}^{\cal F}_E$ for frequencies on one side of the mass gap and less than $\mathcal{A}^{\cal F}_E$ on the other side of the mass gap.

Now we numerically solve \eref{mainEq1}, viewing it as an ``initial'' value problem. First, let us clarifiy its ``initial'' conditions. The $\chi_\pm$ solutions should be regular near the origin $r=0$, which requires $\chi_\pm(\omega, r \to 0) \to F_{\pm}$. Also, thanks to the linearity of \eref{mainEq1}, if $\left(\chi_{+}, \chi_{-}\right)$ is a solution, then $\left(\zeta\chi_{+}, \zeta^*\tilde \eta_{-}\right)$ is also a solution, $\zeta$ being a complex constant. Therefore, we can fix the overall scale of the solutions by scaling (say) $F_{+}$ to unity, which gives us the following ``initial'' condition as $r \rightarrow 0$ :
\be
\label{F1modes}
\chi_{+}(r= 0)=1,~~ \chi_{-}(r=0)=F_{-}.
\ee
The system can be solved with standard ODE solvers in {\tt Matlab}, for example.

Figure \ref{fig:rho1D} and Figure \ref{fig:F1D} respectively depict how the energy amplification factor $\mathcal{A}^{\rho}_E$ and $\mathcal{A}^{\cal F}_E$ change with the frequency $w$ when the ingoing waves only have the $\eta_+$ mode. We have numerically confirmed the superradiant amplification criteria for both of the two definitions with the total and flux scalar energy. 
In these figures, we have chosen a small $h$, compared to $g$, and reducing $h$ further (say $h=0$) only changes the plots very slightly. 

For the case of a single ingoing mode, we see that the two definitions of energy amplification differ significantly near the mass gap. As mentioned, this is due to the fact that the two modes have different group velocities. Compared to the energy flux at infinity, the total energy in a spherical shell gets enhanced on one side of the mass gap but suppressed on the other side, which is an interesting feature for a system with multiple frequency modes. If the shell-energy in a far-away region is more than the energy flux at infinity, it means that, apart from the energy radiated to infinity, there is also energy localized in far-away regions, as a result of the waves scattering on the boson star. It seems that this kind of energy could also be harvested with appropriate means.   

\begin{center}
\begin{figure}
    \includegraphics[width=6.2in]{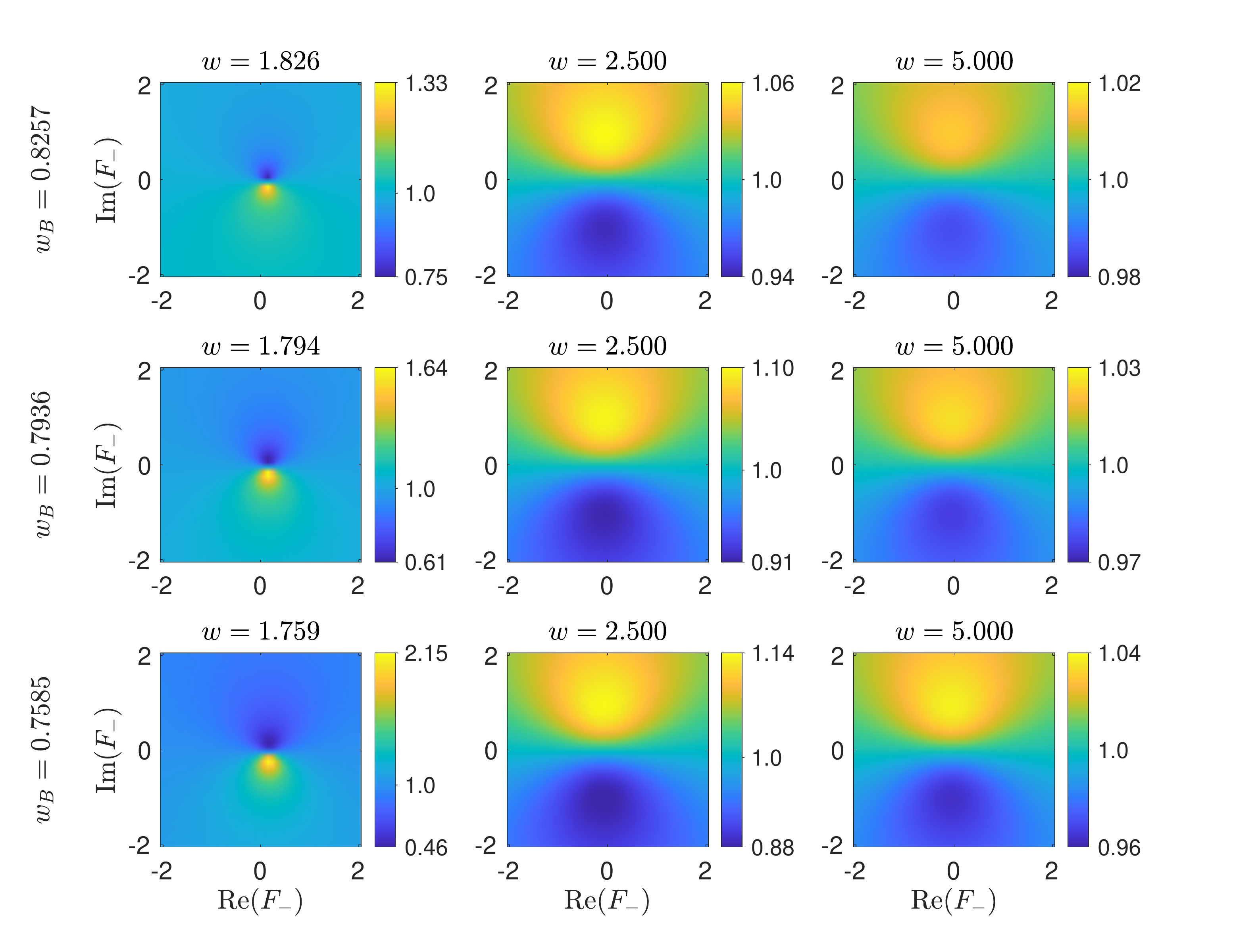}
    \caption{Energy amplification factor $\mathcal{A}^{\rho}_E$ (defined with the total scalar energy) for various $w_B$ and $w$ with both the $+$ and $-$ ingoing modes. The potential parameters are $g=-10$ and $h=0.5$.}
    \label{fig:rho2D}
\end{figure}
\end{center}

\begin{center}
\begin{figure}
    \includegraphics[width=6.2in]{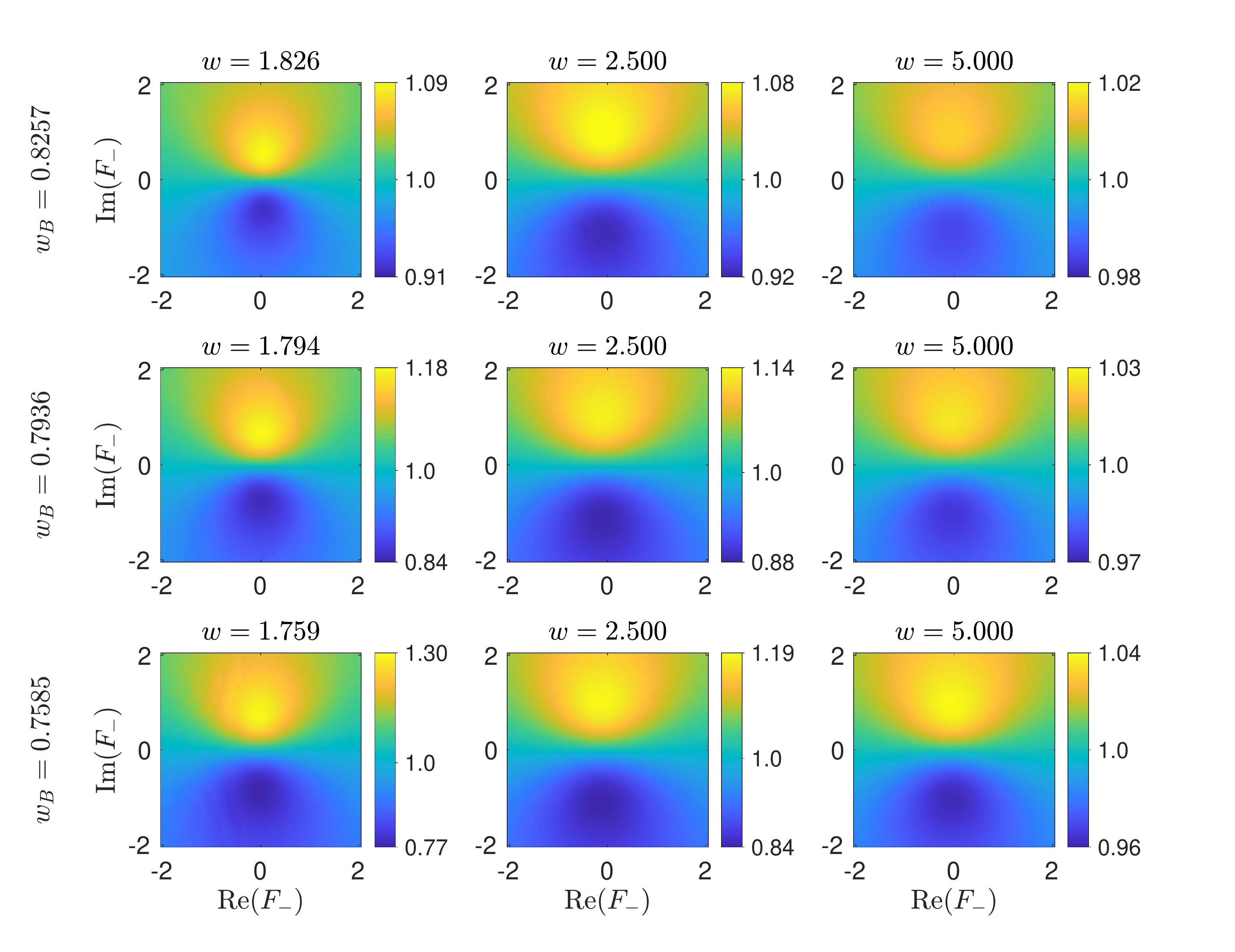}
    \caption{Energy amplification factor $\mathcal{A}^{\cal F}_E$ (defined with the flux scalar energy) for various $w_B$ and $w$ with both $+$ and $-$ ingoing modes. The potential parameters are $g=-10$ and $h=0.5$.}
    \label{fig:F2D}
\end{figure}
\end{center}

Generically, there can be two ingoing modes. In Figure \ref{fig:rho2D} and Figure \ref{fig:F2D}, we plot the energy amplification factor $\mathcal{A}^{\rho}_E$ and $\mathcal{A}^{\cal F}_E$ respectively for various $w_B$ and $w$ with both the $+$ and $-$ ingoing modes. The two modes are parameterized by the complex $F_-$ parameter, which characterizes the behaviors of the modes near the origin. We see that, compared to the single ingoing mode case, the maximum energy amplification factors can be enhanced for certain combinations of the two ingoing modes.

\onecolumngrid

\twocolumngrid

\onecolumngrid

\twocolumngrid

\section{Summary}

\label{sec:summary}

In Ref \cite{Saffin:2022tub}, it has been found that waves scattering on a Q-ball can extract energy from the Q-ball. Due to the similarity between Q-balls and boson stars, it has been postulated that the same energy extraction mechanism works for a boson star. In this paper, we have confirmed this conjecture for the case of relativistic boson stars. The possibility of a similar energy amplification mechanism for a Newtonian boson star has been pointed out by \cite{cardosoSub}. 

For an interacting scalar field, there are generically two coupled modes in the problem of waves scattering  on a boson star. (If one prepares one ingoing mode, generically there will be two outgoing modes.) It can be shown that the particle number of the outgoing waves is exactly the same as the ingoing ones, so energy amplification is possible thanks to the redistribution of particles between the two modes in the scattering. We have shown that energy amplification is possible with only one ingoing mode (and two outgoing modes), but the maximal energy amplification is typically achieved with two ingoing modes. We have computed two ways of defining the energy amplification, one with the total scalar energy and another with the scalar energy flux at infinity. The two definitions differ from each other near the mass gap in the frequency band of the scattering waves. 

We would like to emphasize that, apart from the coherent internal rotation, the form of the scalar potential is also important for the energy extraction here. For Q-balls, only a restricted class of potentials allows for them to form in the first place, but due to the gravitational attraction boson stars can form for more generic classes of scalar potentials. Nevertheless, a preliminary survey of the potentials seems to suggest that sizable energy amplifications can be achieved when the potential is of the Q-ball type. Therefore, internal time-periodic oscillations are only a necessary condition for the amplification mechanism to work, and appropriate interactions are needed to re-distribute the energy among different modes.   

We have referred to the energy extraction mechanism as superradiance or superradiant amplification, although the energy of an individual mode never gets enhanced in the scattering. The term superradiance, as coined by Dicke, originally meant an enhancement of energy in a physical process, and Dicke superradiance is still in active research nowadays (see for example \cite{DickeSuperNat}). For the Q-ball system, since it contains two inseparable modes, it is natural to count all the incoming and outgoing modes to compute the amplification factors. Also, a boson star is a Bose-Einstein condensate within which particles are tightly entangled and oscillates coherently, which is similar to Dicke’s original scenario where superradiance is made possible by a medium of coherent atoms. It is in these contexts that we use the term superradiance, not requiring the energy of an individual mode to be enhanced in the scattering as a necessary qualifier.

We have focused on the simplest case of spherically symmetric boson stars, which are rotating in the (scalar) internal space but not in the real space. A natural extension of this work is to consider rotating boson stars and to investigate how the spatial rotation affect the boson star superradiance \cite{zhangzhou}. Of course, it remains to be shown whether a boson star is subject to superradiant instabilities, which needs to satisfy a stronger criterion than the existence of superradiant amplification that we have demonstrated. Or, one may ask whether there are extra mechanisms to turn the energy amplification into superradiant instabilities. We leave these for future work.

\acknowledgments

We would like to thank Fu-Ming Chang and Guo-Dong Zhang for helpful discusssions. SYZ acknowledges support from the Fundamental Research Funds for the Central Universities under grant No.~WK2030000036, from the National Natural Science Foundation of China under grant No.~12075233 and 12247103, and from the National Key R\&D Program of China under grant No. 2022YFC220010. The work of PMS was funded by STFC Consolidated Grant Number ST/T000732/1.

\bibliography{refs}

\begin{thebibliography}{92}%
\makeatletter
\providecommand \@ifxundefined [1]{%
 \@ifx{#1\undefined}
}%
\providecommand \@ifnum [1]{%
 \ifnum #1\expandafter \@firstoftwo
 \else \expandafter \@secondoftwo
 \fi
}%
\providecommand \@ifx [1]{%
 \ifx #1\expandafter \@firstoftwo
 \else \expandafter \@secondoftwo
 \fi
}%
\providecommand \natexlab [1]{#1}%
\providecommand \enquote  [1]{``#1''}%
\providecommand \bibnamefont  [1]{#1}%
\providecommand \bibfnamefont [1]{#1}%
\providecommand \citenamefont [1]{#1}%
\providecommand \href@noop [0]{\@secondoftwo}%
\providecommand \href [0]{\begingroup \@sanitize@url \@href}%
\providecommand \@href[1]{\@@startlink{#1}\@@href}%
\providecommand \@@href[1]{\endgroup#1\@@endlink}%
\providecommand \@sanitize@url [0]{\catcode `\\12\catcode `\$12\catcode
  `\&12\catcode `\#12\catcode `\^12\catcode `\_12\catcode `\%12\relax}%
\providecommand \@@startlink[1]{}%
\providecommand \@@endlink[0]{}%
\providecommand \url  [0]{\begingroup\@sanitize@url \@url }%
\providecommand \@url [1]{\endgroup\@href {#1}{\urlprefix }}%
\providecommand \urlprefix  [0]{URL }%
\providecommand \Eprint [0]{\href }%
\providecommand \doibase [0]{http://dx.doi.org/}%
\providecommand \selectlanguage [0]{\@gobble}%
\providecommand \bibinfo  [0]{\@secondoftwo}%
\providecommand \bibfield  [0]{\@secondoftwo}%
\providecommand \translation [1]{[#1]}%
\providecommand \BibitemOpen [0]{}%
\providecommand \bibitemStop [0]{}%
\providecommand \bibitemNoStop [0]{.\EOS\space}%
\providecommand \EOS [0]{\spacefactor3000\relax}%
\providecommand \BibitemShut  [1]{\csname bibitem#1\endcsname}%
\let\auto@bib@innerbib\@empty
\bibitem [{\citenamefont {Kaup}(1968)}]{Kaup:1968zz}%
  \BibitemOpen
  \bibfield  {author} {\bibinfo {author} {\bibfnamefont {David~J.}\
  \bibnamefont {Kaup}},\ }\bibfield  {title} {\enquote {\bibinfo {title}
  {{Klein-Gordon Geon}},}\ }\href {\doibase 10.1103/PhysRev.172.1331}
  {\bibfield  {journal} {\bibinfo  {journal} {Phys. Rev.}\ }\textbf {\bibinfo
  {volume} {172}},\ \bibinfo {pages} {1331--1342} (\bibinfo {year}
  {1968})}\BibitemShut {NoStop}%
\bibitem [{\citenamefont {Ruffini}\ and\ \citenamefont
  {Bonazzola}(1969)}]{Ruffini:1969qy}%
  \BibitemOpen
  \bibfield  {author} {\bibinfo {author} {\bibfnamefont {Remo}\ \bibnamefont
  {Ruffini}}\ and\ \bibinfo {author} {\bibfnamefont {Silvano}\ \bibnamefont
  {Bonazzola}},\ }\bibfield  {title} {\enquote {\bibinfo {title} {{Systems of
  selfgravitating particles in general relativity and the concept of an
  equation of state}},}\ }\href {\doibase 10.1103/PhysRev.187.1767} {\bibfield
  {journal} {\bibinfo  {journal} {Phys. Rev.}\ }\textbf {\bibinfo {volume}
  {187}},\ \bibinfo {pages} {1767--1783} (\bibinfo {year} {1969})}\BibitemShut
  {NoStop}%
\bibitem [{\citenamefont {Colpi}\ \emph {et~al.}(1986)\citenamefont {Colpi},
  \citenamefont {Shapiro},\ and\ \citenamefont {Wasserman}}]{Colpi:1986ye}%
  \BibitemOpen
  \bibfield  {author} {\bibinfo {author} {\bibfnamefont {M.}~\bibnamefont
  {Colpi}}, \bibinfo {author} {\bibfnamefont {S.~L.}\ \bibnamefont {Shapiro}},
  \ and\ \bibinfo {author} {\bibfnamefont {I.}~\bibnamefont {Wasserman}},\
  }\bibfield  {title} {\enquote {\bibinfo {title} {{Boson Stars: Gravitational
  Equilibria of Selfinteracting Scalar Fields}},}\ }\href {\doibase
  10.1103/PhysRevLett.57.2485} {\bibfield  {journal} {\bibinfo  {journal}
  {Phys. Rev. Lett.}\ }\textbf {\bibinfo {volume} {57}},\ \bibinfo {pages}
  {2485--2488} (\bibinfo {year} {1986})}\BibitemShut {NoStop}%
\bibitem [{\citenamefont {Friedberg}\ \emph {et~al.}(1987)\citenamefont
  {Friedberg}, \citenamefont {Lee},\ and\ \citenamefont
  {Pang}}]{Friedberg:1986tq}%
  \BibitemOpen
  \bibfield  {author} {\bibinfo {author} {\bibfnamefont {R.}~\bibnamefont
  {Friedberg}}, \bibinfo {author} {\bibfnamefont {T.~D.}\ \bibnamefont {Lee}},
  \ and\ \bibinfo {author} {\bibfnamefont {Y.}~\bibnamefont {Pang}},\
  }\bibfield  {title} {\enquote {\bibinfo {title} {{Scalar Soliton Stars and
  Black Holes}},}\ }\href {\doibase 10.1103/PhysRevD.35.3658} {\bibfield
  {journal} {\bibinfo  {journal} {Phys. Rev. D}\ }\textbf {\bibinfo {volume}
  {35}},\ \bibinfo {pages} {3658} (\bibinfo {year} {1987})}\BibitemShut
  {NoStop}%
\bibitem [{\citenamefont {Jetzer}\ and\ \citenamefont {van~der
  Bij}(1989)}]{Jetzer:1989av}%
  \BibitemOpen
  \bibfield  {author} {\bibinfo {author} {\bibfnamefont {P.}~\bibnamefont
  {Jetzer}}\ and\ \bibinfo {author} {\bibfnamefont {J.~J.}\ \bibnamefont
  {van~der Bij}},\ }\bibfield  {title} {\enquote {\bibinfo {title} {{CHARGED
  BOSON STARS}},}\ }\href {\doibase 10.1016/0370-2693(89)90941-6} {\bibfield
  {journal} {\bibinfo  {journal} {Phys. Lett. B}\ }\textbf {\bibinfo {volume}
  {227}},\ \bibinfo {pages} {341--346} (\bibinfo {year} {1989})}\BibitemShut
  {NoStop}%
\bibitem [{\citenamefont {Henriques}\ \emph {et~al.}(1989)\citenamefont
  {Henriques}, \citenamefont {Liddle},\ and\ \citenamefont
  {Moorhouse}}]{Henriques:1989ar}%
  \BibitemOpen
  \bibfield  {author} {\bibinfo {author} {\bibfnamefont {A.~B.}\ \bibnamefont
  {Henriques}}, \bibinfo {author} {\bibfnamefont {Andrew~R.}\ \bibnamefont
  {Liddle}}, \ and\ \bibinfo {author} {\bibfnamefont {R.~G.}\ \bibnamefont
  {Moorhouse}},\ }\bibfield  {title} {\enquote {\bibinfo {title} {{COMBINED
  BOSON - FERMION STARS}},}\ }\href {\doibase 10.1016/0370-2693(89)90623-0}
  {\bibfield  {journal} {\bibinfo  {journal} {Phys. Lett. B}\ }\textbf
  {\bibinfo {volume} {233}},\ \bibinfo {pages} {99} (\bibinfo {year}
  {1989})}\BibitemShut {NoStop}%
\bibitem [{\citenamefont {Seidel}\ and\ \citenamefont
  {Suen}(1991)}]{Seidel:1991zh}%
  \BibitemOpen
  \bibfield  {author} {\bibinfo {author} {\bibfnamefont {E.}~\bibnamefont
  {Seidel}}\ and\ \bibinfo {author} {\bibfnamefont {W.~M.}\ \bibnamefont
  {Suen}},\ }\bibfield  {title} {\enquote {\bibinfo {title} {{Oscillating
  soliton stars}},}\ }\href {\doibase 10.1103/PhysRevLett.66.1659} {\bibfield
  {journal} {\bibinfo  {journal} {Phys. Rev. Lett.}\ }\textbf {\bibinfo
  {volume} {66}},\ \bibinfo {pages} {1659--1662} (\bibinfo {year}
  {1991})}\BibitemShut {NoStop}%
\bibitem [{\citenamefont {Silveira}\ and\ \citenamefont
  {de~Sousa}(1995)}]{Silveira:1995dh}%
  \BibitemOpen
  \bibfield  {author} {\bibinfo {author} {\bibfnamefont {Vanda}\ \bibnamefont
  {Silveira}}\ and\ \bibinfo {author} {\bibfnamefont {Claudio M.~G.}\
  \bibnamefont {de~Sousa}},\ }\bibfield  {title} {\enquote {\bibinfo {title}
  {{Boson star rotation: A Newtonian approximation}},}\ }\href {\doibase
  10.1103/PhysRevD.52.5724} {\bibfield  {journal} {\bibinfo  {journal} {Phys.
  Rev. D}\ }\textbf {\bibinfo {volume} {52}},\ \bibinfo {pages} {5724--5728}
  (\bibinfo {year} {1995})},\ \Eprint {http://arxiv.org/abs/astro-ph/9508034}
  {arXiv:astro-ph/9508034} \BibitemShut {NoStop}%
\bibitem [{\citenamefont {Schunck}\ and\ \citenamefont
  {Mielke}(1998)}]{Schunck:1996he}%
  \BibitemOpen
  \bibfield  {author} {\bibinfo {author} {\bibfnamefont {France~E.}\
  \bibnamefont {Schunck}}\ and\ \bibinfo {author} {\bibfnamefont {Eckehard~W.}\
  \bibnamefont {Mielke}},\ }\bibfield  {title} {\enquote {\bibinfo {title}
  {{Rotating boson star as an effective mass torus in general relativity}},}\
  }\href {\doibase 10.1016/S0375-9601(98)00778-6} {\bibfield  {journal}
  {\bibinfo  {journal} {Phys. Lett. A}\ }\textbf {\bibinfo {volume} {249}},\
  \bibinfo {pages} {389--394} (\bibinfo {year} {1998})}\BibitemShut {NoStop}%
\bibitem [{\citenamefont {Yoshida}\ and\ \citenamefont
  {Eriguchi}(1997)}]{Yoshida:1997qf}%
  \BibitemOpen
  \bibfield  {author} {\bibinfo {author} {\bibfnamefont {Shijun}\ \bibnamefont
  {Yoshida}}\ and\ \bibinfo {author} {\bibfnamefont {Yoshiharu}\ \bibnamefont
  {Eriguchi}},\ }\bibfield  {title} {\enquote {\bibinfo {title} {{Rotating
  boson stars in general relativity}},}\ }\href {\doibase
  10.1103/PhysRevD.56.762} {\bibfield  {journal} {\bibinfo  {journal} {Phys.
  Rev. D}\ }\textbf {\bibinfo {volume} {56}},\ \bibinfo {pages} {762--771}
  (\bibinfo {year} {1997})}\BibitemShut {NoStop}%
\bibitem [{\citenamefont {Schunck}\ and\ \citenamefont
  {Torres}(2000)}]{Schunck:1999zu}%
  \BibitemOpen
  \bibfield  {author} {\bibinfo {author} {\bibfnamefont {Franz~E.}\
  \bibnamefont {Schunck}}\ and\ \bibinfo {author} {\bibfnamefont {Diego~F.}\
  \bibnamefont {Torres}},\ }\bibfield  {title} {\enquote {\bibinfo {title}
  {{Boson stars with generic selfinteractions}},}\ }\href {\doibase
  10.1142/S0218271800000608} {\bibfield  {journal} {\bibinfo  {journal} {Int.
  J. Mod. Phys. D}\ }\textbf {\bibinfo {volume} {9}},\ \bibinfo {pages}
  {601--618} (\bibinfo {year} {2000})},\ \Eprint
  {http://arxiv.org/abs/gr-qc/9911038} {arXiv:gr-qc/9911038} \BibitemShut
  {NoStop}%
\bibitem [{\citenamefont {Bernal}\ \emph {et~al.}(2010)\citenamefont {Bernal},
  \citenamefont {Barranco}, \citenamefont {Alic},\ and\ \citenamefont
  {Palenzuela}}]{Bernal:2009zy}%
  \BibitemOpen
  \bibfield  {author} {\bibinfo {author} {\bibfnamefont {Argelia}\ \bibnamefont
  {Bernal}}, \bibinfo {author} {\bibfnamefont {Juan}\ \bibnamefont {Barranco}},
  \bibinfo {author} {\bibfnamefont {Daniela}\ \bibnamefont {Alic}}, \ and\
  \bibinfo {author} {\bibfnamefont {Carlos}\ \bibnamefont {Palenzuela}},\
  }\bibfield  {title} {\enquote {\bibinfo {title} {{Multi-state Boson
  Stars}},}\ }\href {\doibase 10.1103/PhysRevD.81.044031} {\bibfield  {journal}
  {\bibinfo  {journal} {Phys. Rev. D}\ }\textbf {\bibinfo {volume} {81}},\
  \bibinfo {pages} {044031} (\bibinfo {year} {2010})},\ \Eprint
  {http://arxiv.org/abs/0908.2435} {arXiv:0908.2435 [gr-qc]} \BibitemShut
  {NoStop}%
\bibitem [{\citenamefont {Herdeiro}\ \emph {et~al.}(2019)\citenamefont
  {Herdeiro}, \citenamefont {Perapechka}, \citenamefont {Radu},\ and\
  \citenamefont {Shnir}}]{Herdeiro:2019mbz}%
  \BibitemOpen
  \bibfield  {author} {\bibinfo {author} {\bibfnamefont {C.}~\bibnamefont
  {Herdeiro}}, \bibinfo {author} {\bibfnamefont {I.}~\bibnamefont
  {Perapechka}}, \bibinfo {author} {\bibfnamefont {E.}~\bibnamefont {Radu}}, \
  and\ \bibinfo {author} {\bibfnamefont {Ya.}\ \bibnamefont {Shnir}},\
  }\bibfield  {title} {\enquote {\bibinfo {title} {{Asymptotically flat
  spinning scalar, Dirac and Proca stars}},}\ }\href {\doibase
  10.1016/j.physletb.2019.134845} {\bibfield  {journal} {\bibinfo  {journal}
  {Phys. Lett. B}\ }\textbf {\bibinfo {volume} {797}},\ \bibinfo {pages}
  {134845} (\bibinfo {year} {2019})},\ \Eprint
  {http://arxiv.org/abs/1906.05386} {arXiv:1906.05386 [gr-qc]} \BibitemShut
  {NoStop}%
\bibitem [{\citenamefont {Liebling}\ and\ \citenamefont
  {Palenzuela}(2012)}]{Liebling:2012fv}%
  \BibitemOpen
  \bibfield  {author} {\bibinfo {author} {\bibfnamefont {Steven~L.}\
  \bibnamefont {Liebling}}\ and\ \bibinfo {author} {\bibfnamefont {Carlos}\
  \bibnamefont {Palenzuela}},\ }\bibfield  {title} {\enquote {\bibinfo {title}
  {{Dynamical boson stars}},}\ }\href {\doibase 10.1007/s41114-023-00043-4}
  {\bibfield  {journal} {\bibinfo  {journal} {Living Rev. Rel.}\ }\textbf
  {\bibinfo {volume} {15}},\ \bibinfo {pages} {6} (\bibinfo {year} {2012})},\
  \Eprint {http://arxiv.org/abs/1202.5809} {arXiv:1202.5809 [gr-qc]}
  \BibitemShut {NoStop}%
\bibitem [{\citenamefont {Visinelli}(2021)}]{Visinelli:2021uve}%
  \BibitemOpen
  \bibfield  {author} {\bibinfo {author} {\bibfnamefont {Luca}\ \bibnamefont
  {Visinelli}},\ }\bibfield  {title} {\enquote {\bibinfo {title} {{Boson stars
  and oscillatons: A review}},}\ }\href {\doibase 10.1142/S0218271821300068}
  {\bibfield  {journal} {\bibinfo  {journal} {Int. J. Mod. Phys. D}\ }\textbf
  {\bibinfo {volume} {30}},\ \bibinfo {pages} {2130006} (\bibinfo {year}
  {2021})},\ \Eprint {http://arxiv.org/abs/2109.05481} {arXiv:2109.05481
  [gr-qc]} \BibitemShut {NoStop}%
\bibitem [{\citenamefont {Lee}\ and\ \citenamefont {Koh}(1996)}]{Lee:1995af}%
  \BibitemOpen
  \bibfield  {author} {\bibinfo {author} {\bibfnamefont {Jae-weon}\
  \bibnamefont {Lee}}\ and\ \bibinfo {author} {\bibfnamefont {In-gyu}\
  \bibnamefont {Koh}},\ }\bibfield  {title} {\enquote {\bibinfo {title}
  {{Galactic halos as boson stars}},}\ }\href {\doibase
  10.1103/PhysRevD.53.2236} {\bibfield  {journal} {\bibinfo  {journal} {Phys.
  Rev. D}\ }\textbf {\bibinfo {volume} {53}},\ \bibinfo {pages} {2236--2239}
  (\bibinfo {year} {1996})},\ \Eprint {http://arxiv.org/abs/hep-ph/9507385}
  {arXiv:hep-ph/9507385} \BibitemShut {NoStop}%
\bibitem [{\citenamefont {Chavanis}(2011)}]{Chavanis:2011zi}%
  \BibitemOpen
  \bibfield  {author} {\bibinfo {author} {\bibfnamefont {Pierre-Henri}\
  \bibnamefont {Chavanis}},\ }\bibfield  {title} {\enquote {\bibinfo {title}
  {{Mass-radius relation of Newtonian self-gravitating Bose-Einstein
  condensates with short-range interactions: I. Analytical results}},}\ }\href
  {\doibase 10.1103/PhysRevD.84.043531} {\bibfield  {journal} {\bibinfo
  {journal} {Phys. Rev. D}\ }\textbf {\bibinfo {volume} {84}},\ \bibinfo
  {pages} {043531} (\bibinfo {year} {2011})},\ \Eprint
  {http://arxiv.org/abs/1103.2050} {arXiv:1103.2050 [astro-ph.CO]} \BibitemShut
  {NoStop}%
\bibitem [{\citenamefont {Rindler-Daller}\ and\ \citenamefont
  {Shapiro}(2012)}]{Rindler-Daller:2011afd}%
  \BibitemOpen
  \bibfield  {author} {\bibinfo {author} {\bibfnamefont {Tanja}\ \bibnamefont
  {Rindler-Daller}}\ and\ \bibinfo {author} {\bibfnamefont {Paul~R.}\
  \bibnamefont {Shapiro}},\ }\bibfield  {title} {\enquote {\bibinfo {title}
  {{Angular Momentum and Vortex Formation in Bose-Einstein-Condensed Cold Dark
  Matter Haloes}},}\ }\href {\doibase 10.1111/j.1365-2966.2012.20588.x}
  {\bibfield  {journal} {\bibinfo  {journal} {Mon. Not. Roy. Astron. Soc.}\
  }\textbf {\bibinfo {volume} {422}},\ \bibinfo {pages} {135--161} (\bibinfo
  {year} {2012})},\ \Eprint {http://arxiv.org/abs/1106.1256} {arXiv:1106.1256
  [astro-ph.CO]} \BibitemShut {NoStop}%
\bibitem [{\citenamefont {Eby}\ \emph {et~al.}(2016)\citenamefont {Eby},
  \citenamefont {Kouvaris}, \citenamefont {Nielsen},\ and\ \citenamefont
  {Wijewardhana}}]{Eby:2015hsq}%
  \BibitemOpen
  \bibfield  {author} {\bibinfo {author} {\bibfnamefont {Joshua}\ \bibnamefont
  {Eby}}, \bibinfo {author} {\bibfnamefont {Chris}\ \bibnamefont {Kouvaris}},
  \bibinfo {author} {\bibfnamefont {Niklas~Gr\o{}nlund}\ \bibnamefont
  {Nielsen}}, \ and\ \bibinfo {author} {\bibfnamefont {L.~C.~R.}\ \bibnamefont
  {Wijewardhana}},\ }\bibfield  {title} {\enquote {\bibinfo {title} {{Boson
  Stars from Self-Interacting Dark Matter}},}\ }\href {\doibase
  10.1007/JHEP02(2016)028} {\bibfield  {journal} {\bibinfo  {journal} {JHEP}\
  }\textbf {\bibinfo {volume} {02}},\ \bibinfo {pages} {028} (\bibinfo {year}
  {2016})},\ \Eprint {http://arxiv.org/abs/1511.04474} {arXiv:1511.04474
  [hep-ph]} \BibitemShut {NoStop}%
\bibitem [{\citenamefont {March-Russell}\ and\ \citenamefont
  {Rosa}(2022)}]{March-Russell:2022zll}%
  \BibitemOpen
  \bibfield  {author} {\bibinfo {author} {\bibfnamefont {John}\ \bibnamefont
  {March-Russell}}\ and\ \bibinfo {author} {\bibfnamefont {Jo\~ao~G.}\
  \bibnamefont {Rosa}},\ }\bibfield  {title} {\enquote {\bibinfo {title}
  {{Micro-Bose/Proca dark matter stars from black hole superradiance}},}\
  }\href@noop {} {\  (\bibinfo {year} {2022})},\ \Eprint
  {http://arxiv.org/abs/2205.15277} {arXiv:2205.15277 [gr-qc]} \BibitemShut
  {NoStop}%
\bibitem [{\citenamefont {Navarro-Boullosa}\ \emph {et~al.}(2023)\citenamefont
  {Navarro-Boullosa}, \citenamefont {Bernal},\ and\ \citenamefont
  {Vazquez}}]{Navarro-Boullosa:2023bya}%
  \BibitemOpen
  \bibfield  {author} {\bibinfo {author} {\bibfnamefont {Atalia}\ \bibnamefont
  {Navarro-Boullosa}}, \bibinfo {author} {\bibfnamefont {Argelia}\ \bibnamefont
  {Bernal}}, \ and\ \bibinfo {author} {\bibfnamefont {J.~Alberto}\ \bibnamefont
  {Vazquez}},\ }\bibfield  {title} {\enquote {\bibinfo {title} {{Bayesian
  analysis for rotational curves with $\ell$-boson stars as a dark matter
  component}},}\ }\href@noop {} {\  (\bibinfo {year} {2023})},\ \Eprint
  {http://arxiv.org/abs/2305.01127} {arXiv:2305.01127 [astro-ph.CO]}
  \BibitemShut {NoStop}%
\bibitem [{\citenamefont {Torres}\ \emph {et~al.}(2000)\citenamefont {Torres},
  \citenamefont {Capozziello},\ and\ \citenamefont {Lambiase}}]{Torres:2000dw}%
  \BibitemOpen
  \bibfield  {author} {\bibinfo {author} {\bibfnamefont {Diego~F.}\
  \bibnamefont {Torres}}, \bibinfo {author} {\bibfnamefont {S.}~\bibnamefont
  {Capozziello}}, \ and\ \bibinfo {author} {\bibfnamefont {G.}~\bibnamefont
  {Lambiase}},\ }\bibfield  {title} {\enquote {\bibinfo {title} {{A
  Supermassive scalar star at the galactic center?}}}\ }\href {\doibase
  10.1103/PhysRevD.62.104012} {\bibfield  {journal} {\bibinfo  {journal} {Phys.
  Rev. D}\ }\textbf {\bibinfo {volume} {62}},\ \bibinfo {pages} {104012}
  (\bibinfo {year} {2000})},\ \Eprint {http://arxiv.org/abs/astro-ph/0004064}
  {arXiv:astro-ph/0004064} \BibitemShut {NoStop}%
\bibitem [{\citenamefont {Guzman}\ and\ \citenamefont
  {Rueda-Becerril}(2009)}]{Guzman:2009zz}%
  \BibitemOpen
  \bibfield  {author} {\bibinfo {author} {\bibfnamefont {F.~S.}\ \bibnamefont
  {Guzman}}\ and\ \bibinfo {author} {\bibfnamefont {J.~M.}\ \bibnamefont
  {Rueda-Becerril}},\ }\bibfield  {title} {\enquote {\bibinfo {title}
  {{Spherical boson stars as black hole mimickers}},}\ }\href {\doibase
  10.1103/PhysRevD.80.084023} {\bibfield  {journal} {\bibinfo  {journal} {Phys.
  Rev. D}\ }\textbf {\bibinfo {volume} {80}},\ \bibinfo {pages} {084023}
  (\bibinfo {year} {2009})},\ \Eprint {http://arxiv.org/abs/1009.1250}
  {arXiv:1009.1250 [astro-ph.HE]} \BibitemShut {NoStop}%
\bibitem [{\citenamefont {Olivares}\ \emph {et~al.}(2020)\citenamefont
  {Olivares}, \citenamefont {Younsi}, \citenamefont {Fromm}, \citenamefont
  {De~Laurentis}, \citenamefont {Porth}, \citenamefont {Mizuno}, \citenamefont
  {Falcke}, \citenamefont {Kramer},\ and\ \citenamefont
  {Rezzolla}}]{Olivares:2018abq}%
  \BibitemOpen
  \bibfield  {author} {\bibinfo {author} {\bibfnamefont {Hector}\ \bibnamefont
  {Olivares}}, \bibinfo {author} {\bibfnamefont {Ziri}\ \bibnamefont {Younsi}},
  \bibinfo {author} {\bibfnamefont {Christian~M.}\ \bibnamefont {Fromm}},
  \bibinfo {author} {\bibfnamefont {Mariafelicia}\ \bibnamefont
  {De~Laurentis}}, \bibinfo {author} {\bibfnamefont {Oliver}\ \bibnamefont
  {Porth}}, \bibinfo {author} {\bibfnamefont {Yosuke}\ \bibnamefont {Mizuno}},
  \bibinfo {author} {\bibfnamefont {Heino}\ \bibnamefont {Falcke}}, \bibinfo
  {author} {\bibfnamefont {Michael}\ \bibnamefont {Kramer}}, \ and\ \bibinfo
  {author} {\bibfnamefont {Luciano}\ \bibnamefont {Rezzolla}},\ }\bibfield
  {title} {\enquote {\bibinfo {title} {{How to tell an accreting boson star
  from a black hole}},}\ }\href {\doibase 10.1093/mnras/staa1878} {\bibfield
  {journal} {\bibinfo  {journal} {Mon. Not. Roy. Astron. Soc.}\ }\textbf
  {\bibinfo {volume} {497}},\ \bibinfo {pages} {521--535} (\bibinfo {year}
  {2020})},\ \Eprint {http://arxiv.org/abs/1809.08682} {arXiv:1809.08682
  [gr-qc]} \BibitemShut {NoStop}%
\bibitem [{\citenamefont {Herdeiro}\ \emph {et~al.}(2021)\citenamefont
  {Herdeiro}, \citenamefont {Pombo}, \citenamefont {Radu}, \citenamefont
  {Cunha},\ and\ \citenamefont {Sanchis-Gual}}]{Herdeiro:2021lwl}%
  \BibitemOpen
  \bibfield  {author} {\bibinfo {author} {\bibfnamefont {Carlos A.~R.}\
  \bibnamefont {Herdeiro}}, \bibinfo {author} {\bibfnamefont {Alexandre~M.}\
  \bibnamefont {Pombo}}, \bibinfo {author} {\bibfnamefont {Eugen}\ \bibnamefont
  {Radu}}, \bibinfo {author} {\bibfnamefont {Pedro V.~P.}\ \bibnamefont
  {Cunha}}, \ and\ \bibinfo {author} {\bibfnamefont {Nicolas}\ \bibnamefont
  {Sanchis-Gual}},\ }\bibfield  {title} {\enquote {\bibinfo {title} {{The
  imitation game: Proca stars that can mimic the Schwarzschild shadow}},}\
  }\href {\doibase 10.1088/1475-7516/2021/04/051} {\bibfield  {journal}
  {\bibinfo  {journal} {JCAP}\ }\textbf {\bibinfo {volume} {04}},\ \bibinfo
  {pages} {051} (\bibinfo {year} {2021})},\ \Eprint
  {http://arxiv.org/abs/2102.01703} {arXiv:2102.01703 [gr-qc]} \BibitemShut
  {NoStop}%
\bibitem [{\citenamefont {Berti}\ and\ \citenamefont
  {Cardoso}(2006)}]{Berti:2006qt}%
  \BibitemOpen
  \bibfield  {author} {\bibinfo {author} {\bibfnamefont {Emanuele}\
  \bibnamefont {Berti}}\ and\ \bibinfo {author} {\bibfnamefont {Vitor}\
  \bibnamefont {Cardoso}},\ }\bibfield  {title} {\enquote {\bibinfo {title}
  {{Supermassive black holes or boson stars? Hair counting with gravitational
  wave detectors}},}\ }\href {\doibase 10.1142/S0218271806009637} {\bibfield
  {journal} {\bibinfo  {journal} {Int. J. Mod. Phys. D}\ }\textbf {\bibinfo
  {volume} {15}},\ \bibinfo {pages} {2209--2216} (\bibinfo {year} {2006})},\
  \Eprint {http://arxiv.org/abs/gr-qc/0605101} {arXiv:gr-qc/0605101}
  \BibitemShut {NoStop}%
\bibitem [{\citenamefont {Palenzuela}\ \emph {et~al.}(2008)\citenamefont
  {Palenzuela}, \citenamefont {Lehner},\ and\ \citenamefont
  {Liebling}}]{Palenzuela:2007dm}%
  \BibitemOpen
  \bibfield  {author} {\bibinfo {author} {\bibfnamefont {C.}~\bibnamefont
  {Palenzuela}}, \bibinfo {author} {\bibfnamefont {L.}~\bibnamefont {Lehner}},
  \ and\ \bibinfo {author} {\bibfnamefont {Steven~L.}\ \bibnamefont
  {Liebling}},\ }\bibfield  {title} {\enquote {\bibinfo {title} {{Orbital
  Dynamics of Binary Boson Star Systems}},}\ }\href {\doibase
  10.1103/PhysRevD.77.044036} {\bibfield  {journal} {\bibinfo  {journal} {Phys.
  Rev. D}\ }\textbf {\bibinfo {volume} {77}},\ \bibinfo {pages} {044036}
  (\bibinfo {year} {2008})},\ \Eprint {http://arxiv.org/abs/0706.2435}
  {arXiv:0706.2435 [gr-qc]} \BibitemShut {NoStop}%
\bibitem [{\citenamefont {Cardoso}\ \emph {et~al.}(2016)\citenamefont
  {Cardoso}, \citenamefont {Hopper}, \citenamefont {Macedo}, \citenamefont
  {Palenzuela},\ and\ \citenamefont {Pani}}]{Cardoso:2016oxy}%
  \BibitemOpen
  \bibfield  {author} {\bibinfo {author} {\bibfnamefont {Vitor}\ \bibnamefont
  {Cardoso}}, \bibinfo {author} {\bibfnamefont {Seth}\ \bibnamefont {Hopper}},
  \bibinfo {author} {\bibfnamefont {Caio F.~B.}\ \bibnamefont {Macedo}},
  \bibinfo {author} {\bibfnamefont {Carlos}\ \bibnamefont {Palenzuela}}, \ and\
  \bibinfo {author} {\bibfnamefont {Paolo}\ \bibnamefont {Pani}},\ }\bibfield
  {title} {\enquote {\bibinfo {title} {{Gravitational-wave signatures of exotic
  compact objects and of quantum corrections at the horizon scale}},}\ }\href
  {\doibase 10.1103/PhysRevD.94.084031} {\bibfield  {journal} {\bibinfo
  {journal} {Phys. Rev. D}\ }\textbf {\bibinfo {volume} {94}},\ \bibinfo
  {pages} {084031} (\bibinfo {year} {2016})},\ \Eprint
  {http://arxiv.org/abs/1608.08637} {arXiv:1608.08637 [gr-qc]} \BibitemShut
  {NoStop}%
\bibitem [{\citenamefont {Sennett}\ \emph {et~al.}(2017)\citenamefont
  {Sennett}, \citenamefont {Hinderer}, \citenamefont {Steinhoff}, \citenamefont
  {Buonanno},\ and\ \citenamefont {Ossokine}}]{Sennett:2017etc}%
  \BibitemOpen
  \bibfield  {author} {\bibinfo {author} {\bibfnamefont {Noah}\ \bibnamefont
  {Sennett}}, \bibinfo {author} {\bibfnamefont {Tanja}\ \bibnamefont
  {Hinderer}}, \bibinfo {author} {\bibfnamefont {Jan}\ \bibnamefont
  {Steinhoff}}, \bibinfo {author} {\bibfnamefont {Alessandra}\ \bibnamefont
  {Buonanno}}, \ and\ \bibinfo {author} {\bibfnamefont {Serguei}\ \bibnamefont
  {Ossokine}},\ }\bibfield  {title} {\enquote {\bibinfo {title}
  {{Distinguishing Boson Stars from Black Holes and Neutron Stars from Tidal
  Interactions in Inspiraling Binary Systems}},}\ }\href {\doibase
  10.1103/PhysRevD.96.024002} {\bibfield  {journal} {\bibinfo  {journal} {Phys.
  Rev. D}\ }\textbf {\bibinfo {volume} {96}},\ \bibinfo {pages} {024002}
  (\bibinfo {year} {2017})},\ \Eprint {http://arxiv.org/abs/1704.08651}
  {arXiv:1704.08651 [gr-qc]} \BibitemShut {NoStop}%
\bibitem [{\citenamefont {Palenzuela}\ \emph {et~al.}(2017)\citenamefont
  {Palenzuela}, \citenamefont {Pani}, \citenamefont {Bezares}, \citenamefont
  {Cardoso}, \citenamefont {Lehner},\ and\ \citenamefont
  {Liebling}}]{Palenzuela:2017kcg}%
  \BibitemOpen
  \bibfield  {author} {\bibinfo {author} {\bibfnamefont {Carlos}\ \bibnamefont
  {Palenzuela}}, \bibinfo {author} {\bibfnamefont {Paolo}\ \bibnamefont
  {Pani}}, \bibinfo {author} {\bibfnamefont {Miguel}\ \bibnamefont {Bezares}},
  \bibinfo {author} {\bibfnamefont {Vitor}\ \bibnamefont {Cardoso}}, \bibinfo
  {author} {\bibfnamefont {Luis}\ \bibnamefont {Lehner}}, \ and\ \bibinfo
  {author} {\bibfnamefont {Steven}\ \bibnamefont {Liebling}},\ }\bibfield
  {title} {\enquote {\bibinfo {title} {{Gravitational Wave Signatures of Highly
  Compact Boson Star Binaries}},}\ }\href {\doibase 10.1103/PhysRevD.96.104058}
  {\bibfield  {journal} {\bibinfo  {journal} {Phys. Rev. D}\ }\textbf {\bibinfo
  {volume} {96}},\ \bibinfo {pages} {104058} (\bibinfo {year} {2017})},\
  \Eprint {http://arxiv.org/abs/1710.09432} {arXiv:1710.09432 [gr-qc]}
  \BibitemShut {NoStop}%
\bibitem [{\citenamefont {Dietrich}\ \emph {et~al.}(2019)\citenamefont
  {Dietrich}, \citenamefont {Ossokine},\ and\ \citenamefont
  {Clough}}]{Dietrich:2018bvi}%
  \BibitemOpen
  \bibfield  {author} {\bibinfo {author} {\bibfnamefont {Tim}\ \bibnamefont
  {Dietrich}}, \bibinfo {author} {\bibfnamefont {Serguei}\ \bibnamefont
  {Ossokine}}, \ and\ \bibinfo {author} {\bibfnamefont {Katy}\ \bibnamefont
  {Clough}},\ }\bibfield  {title} {\enquote {\bibinfo {title} {{Full 3D
  numerical relativity simulations of neutron star\textendash{}boson star
  collisions with BAM}},}\ }\href {\doibase 10.1088/1361-6382/aaf43e}
  {\bibfield  {journal} {\bibinfo  {journal} {Class. Quant. Grav.}\ }\textbf
  {\bibinfo {volume} {36}},\ \bibinfo {pages} {025002} (\bibinfo {year}
  {2019})},\ \Eprint {http://arxiv.org/abs/1807.06959} {arXiv:1807.06959
  [gr-qc]} \BibitemShut {NoStop}%
\bibitem [{\citenamefont {Bezares}\ and\ \citenamefont
  {Palenzuela}(2018)}]{Bezares:2018qwa}%
  \BibitemOpen
  \bibfield  {author} {\bibinfo {author} {\bibfnamefont {Miguel}\ \bibnamefont
  {Bezares}}\ and\ \bibinfo {author} {\bibfnamefont {Carlos}\ \bibnamefont
  {Palenzuela}},\ }\bibfield  {title} {\enquote {\bibinfo {title}
  {{Gravitational Waves from Dark Boson Star binary mergers}},}\ }\href
  {\doibase 10.1088/1361-6382/aae87c} {\bibfield  {journal} {\bibinfo
  {journal} {Class. Quant. Grav.}\ }\textbf {\bibinfo {volume} {35}},\ \bibinfo
  {pages} {234002} (\bibinfo {year} {2018})},\ \Eprint
  {http://arxiv.org/abs/1808.10732} {arXiv:1808.10732 [gr-qc]} \BibitemShut
  {NoStop}%
\bibitem [{\citenamefont {Croon}\ \emph {et~al.}(2019)\citenamefont {Croon},
  \citenamefont {Fan},\ and\ \citenamefont {Sun}}]{Croon:2018ybs}%
  \BibitemOpen
  \bibfield  {author} {\bibinfo {author} {\bibfnamefont {Djuna}\ \bibnamefont
  {Croon}}, \bibinfo {author} {\bibfnamefont {Jiji}\ \bibnamefont {Fan}}, \
  and\ \bibinfo {author} {\bibfnamefont {Chen}\ \bibnamefont {Sun}},\
  }\bibfield  {title} {\enquote {\bibinfo {title} {{Boson Star from Repulsive
  Light Scalars and Gravitational Waves}},}\ }\href {\doibase
  10.1088/1475-7516/2019/04/008} {\bibfield  {journal} {\bibinfo  {journal}
  {JCAP}\ }\textbf {\bibinfo {volume} {04}},\ \bibinfo {pages} {008} (\bibinfo
  {year} {2019})},\ \Eprint {http://arxiv.org/abs/1810.01420} {arXiv:1810.01420
  [hep-ph]} \BibitemShut {NoStop}%
\bibitem [{\citenamefont {Choi}\ \emph {et~al.}(2019)\citenamefont {Choi},
  \citenamefont {He},\ and\ \citenamefont {Schiappacasse}}]{Choi:2019mva}%
  \BibitemOpen
  \bibfield  {author} {\bibinfo {author} {\bibfnamefont {Gongjun}\ \bibnamefont
  {Choi}}, \bibinfo {author} {\bibfnamefont {Hong-Jian}\ \bibnamefont {He}}, \
  and\ \bibinfo {author} {\bibfnamefont {Enrico~D.}\ \bibnamefont
  {Schiappacasse}},\ }\bibfield  {title} {\enquote {\bibinfo {title} {{Probing
  Dynamics of Boson Stars by Fast Radio Bursts and Gravitational Wave
  Detection}},}\ }\href {\doibase 10.1088/1475-7516/2019/10/043} {\bibfield
  {journal} {\bibinfo  {journal} {JCAP}\ }\textbf {\bibinfo {volume} {10}},\
  \bibinfo {pages} {043} (\bibinfo {year} {2019})},\ \Eprint
  {http://arxiv.org/abs/1906.02094} {arXiv:1906.02094 [astro-ph.CO]}
  \BibitemShut {NoStop}%
\bibitem [{\citenamefont {Bezares}\ \emph {et~al.}(2022)\citenamefont
  {Bezares}, \citenamefont {Bo\v{s}kovi\'c}, \citenamefont {Liebling},
  \citenamefont {Palenzuela}, \citenamefont {Pani},\ and\ \citenamefont
  {Barausse}}]{Bezares:2022obu}%
  \BibitemOpen
  \bibfield  {author} {\bibinfo {author} {\bibfnamefont {Miguel}\ \bibnamefont
  {Bezares}}, \bibinfo {author} {\bibfnamefont {Mateja}\ \bibnamefont
  {Bo\v{s}kovi\'c}}, \bibinfo {author} {\bibfnamefont {Steven}\ \bibnamefont
  {Liebling}}, \bibinfo {author} {\bibfnamefont {Carlos}\ \bibnamefont
  {Palenzuela}}, \bibinfo {author} {\bibfnamefont {Paolo}\ \bibnamefont
  {Pani}}, \ and\ \bibinfo {author} {\bibfnamefont {Enrico}\ \bibnamefont
  {Barausse}},\ }\bibfield  {title} {\enquote {\bibinfo {title} {{Gravitational
  waves and kicks from the merger of unequal mass, highly compact boson
  stars}},}\ }\href {\doibase 10.1103/PhysRevD.105.064067} {\bibfield
  {journal} {\bibinfo  {journal} {Phys. Rev. D}\ }\textbf {\bibinfo {volume}
  {105}},\ \bibinfo {pages} {064067} (\bibinfo {year} {2022})},\ \Eprint
  {http://arxiv.org/abs/2201.06113} {arXiv:2201.06113 [gr-qc]} \BibitemShut
  {NoStop}%
\bibitem [{\citenamefont {Jaramillo}\ \emph {et~al.}(2022)\citenamefont
  {Jaramillo}, \citenamefont {Sanchis-Gual}, \citenamefont {Barranco},
  \citenamefont {Bernal}, \citenamefont {Degollado}, \citenamefont {Herdeiro},
  \citenamefont {Megevand},\ and\ \citenamefont
  {N\'u\~nez}}]{Jaramillo:2022zwg}%
  \BibitemOpen
  \bibfield  {author} {\bibinfo {author} {\bibfnamefont {V\'\i{}ctor}\
  \bibnamefont {Jaramillo}}, \bibinfo {author} {\bibfnamefont {Nicolas}\
  \bibnamefont {Sanchis-Gual}}, \bibinfo {author} {\bibfnamefont {Juan}\
  \bibnamefont {Barranco}}, \bibinfo {author} {\bibfnamefont {Argelia}\
  \bibnamefont {Bernal}}, \bibinfo {author} {\bibfnamefont {Juan~Carlos}\
  \bibnamefont {Degollado}}, \bibinfo {author} {\bibfnamefont {Carlos}\
  \bibnamefont {Herdeiro}}, \bibinfo {author} {\bibfnamefont {Miguel}\
  \bibnamefont {Megevand}}, \ and\ \bibinfo {author} {\bibfnamefont
  {Dar\'\i{}o}\ \bibnamefont {N\'u\~nez}},\ }\bibfield  {title} {\enquote
  {\bibinfo {title} {{Head-on collisions of \ensuremath{\ell}-boson stars}},}\
  }\href {\doibase 10.1103/PhysRevD.105.104057} {\bibfield  {journal} {\bibinfo
   {journal} {Phys. Rev. D}\ }\textbf {\bibinfo {volume} {105}},\ \bibinfo
  {pages} {104057} (\bibinfo {year} {2022})},\ \Eprint
  {http://arxiv.org/abs/2202.00696} {arXiv:2202.00696 [gr-qc]} \BibitemShut
  {NoStop}%
\bibitem [{\citenamefont {Croft}\ \emph {et~al.}(2023)\citenamefont {Croft},
  \citenamefont {Helfer}, \citenamefont {Ge}, \citenamefont {Radia},
  \citenamefont {Evstafyeva}, \citenamefont {Lim}, \citenamefont {Sperhake},\
  and\ \citenamefont {Clough}}]{Croft:2022bxq}%
  \BibitemOpen
  \bibfield  {author} {\bibinfo {author} {\bibfnamefont {Robin}\ \bibnamefont
  {Croft}}, \bibinfo {author} {\bibfnamefont {Thomas}\ \bibnamefont {Helfer}},
  \bibinfo {author} {\bibfnamefont {Bo-Xuan}\ \bibnamefont {Ge}}, \bibinfo
  {author} {\bibfnamefont {Miren}\ \bibnamefont {Radia}}, \bibinfo {author}
  {\bibfnamefont {Tamara}\ \bibnamefont {Evstafyeva}}, \bibinfo {author}
  {\bibfnamefont {Eugene~A.}\ \bibnamefont {Lim}}, \bibinfo {author}
  {\bibfnamefont {Ulrich}\ \bibnamefont {Sperhake}}, \ and\ \bibinfo {author}
  {\bibfnamefont {Katy}\ \bibnamefont {Clough}},\ }\bibfield  {title} {\enquote
  {\bibinfo {title} {{The gravitational afterglow of boson stars}},}\ }\href
  {\doibase 10.1088/1361-6382/acace4} {\bibfield  {journal} {\bibinfo
  {journal} {Class. Quant. Grav.}\ }\textbf {\bibinfo {volume} {40}},\ \bibinfo
  {pages} {065001} (\bibinfo {year} {2023})},\ \Eprint
  {http://arxiv.org/abs/2207.05690} {arXiv:2207.05690 [gr-qc]} \BibitemShut
  {NoStop}%
\bibitem [{\citenamefont {Dicke}(1954)}]{Dicke:1954zz}%
  \BibitemOpen
  \bibfield  {author} {\bibinfo {author} {\bibfnamefont {R.~H.}\ \bibnamefont
  {Dicke}},\ }\bibfield  {title} {\enquote {\bibinfo {title} {{Coherence in
  Spontaneous Radiation Processes}},}\ }\href {\doibase 10.1103/PhysRev.93.99}
  {\bibfield  {journal} {\bibinfo  {journal} {Phys. Rev.}\ }\textbf {\bibinfo
  {volume} {93}},\ \bibinfo {pages} {99--110} (\bibinfo {year}
  {1954})}\BibitemShut {NoStop}%
\bibitem [{\citenamefont {Zel'dovich}({\natexlab{a}})}]{Zeld1}%
  \BibitemOpen
  \bibfield  {author} {\bibinfo {author} {\bibfnamefont {Ya.~B.}\ \bibnamefont
  {Zel'dovich}},\ }\bibfield  {title} {\enquote {\bibinfo {title} {{GENERATION
  OF WAVES BY A ROTATING BODY}},}\ }\href@noop {} {\bibfield  {journal}
  {\bibinfo  {journal} {Zh. Eksp. Teor. Fiz. Pis'ma {\bf 14}, 270 (1971) [JETP
  Letters {\bf 14}, 180 (1971)]}\ } ({\natexlab{a}})}\BibitemShut {NoStop}%
\bibitem [{\citenamefont {Zel'dovich}({\natexlab{b}})}]{Zeld2}%
  \BibitemOpen
  \bibfield  {author} {\bibinfo {author} {\bibfnamefont {Ya.~B.}\ \bibnamefont
  {Zel'dovich}},\ }\bibfield  {title} {\enquote {\bibinfo {title}
  {{Amplification of Cylindrical Electromagnetic Waves Reflected from a
  Rotating Body}},}\ }\href@noop {} {\bibfield  {journal} {\bibinfo  {journal}
  {Zh. Eksp. Teor. Fiz. {\bf 62}, 2076 (1971) [JETP {\bf 35}, 1085 (1971)]}\ }
  ({\natexlab{b}})}\BibitemShut {NoStop}%
\bibitem [{\citenamefont {Misner}(1972)}]{Misner:1972kx}%
  \BibitemOpen
  \bibfield  {author} {\bibinfo {author} {\bibfnamefont {Charles~W.}\
  \bibnamefont {Misner}},\ }\bibfield  {title} {\enquote {\bibinfo {title}
  {{Interpretation of gravitational-wave observations}},}\ }\href {\doibase
  10.1103/PhysRevLett.28.994} {\bibfield  {journal} {\bibinfo  {journal} {Phys.
  Rev. Lett.}\ }\textbf {\bibinfo {volume} {28}},\ \bibinfo {pages} {994--997}
  (\bibinfo {year} {1972})}\BibitemShut {NoStop}%
\bibitem [{\citenamefont {Cardoso}\ and\ \citenamefont
  {Dias}(2004)}]{Cardoso:2004hs}%
  \BibitemOpen
  \bibfield  {author} {\bibinfo {author} {\bibfnamefont {Vitor}\ \bibnamefont
  {Cardoso}}\ and\ \bibinfo {author} {\bibfnamefont {Oscar J.~C.}\ \bibnamefont
  {Dias}},\ }\bibfield  {title} {\enquote {\bibinfo {title} {{Small
  Kerr-anti-de Sitter black holes are unstable}},}\ }\href {\doibase
  10.1103/PhysRevD.70.084011} {\bibfield  {journal} {\bibinfo  {journal} {Phys.
  Rev. D}\ }\textbf {\bibinfo {volume} {70}},\ \bibinfo {pages} {084011}
  (\bibinfo {year} {2004})},\ \Eprint {http://arxiv.org/abs/hep-th/0405006}
  {arXiv:hep-th/0405006} \BibitemShut {NoStop}%
\bibitem [{\citenamefont {Dolan}(2007)}]{Dolan:2007mj}%
  \BibitemOpen
  \bibfield  {author} {\bibinfo {author} {\bibfnamefont {Sam~R.}\ \bibnamefont
  {Dolan}},\ }\bibfield  {title} {\enquote {\bibinfo {title} {{Instability of
  the massive Klein-Gordon field on the Kerr spacetime}},}\ }\href {\doibase
  10.1103/PhysRevD.76.084001} {\bibfield  {journal} {\bibinfo  {journal} {Phys.
  Rev. D}\ }\textbf {\bibinfo {volume} {76}},\ \bibinfo {pages} {084001}
  (\bibinfo {year} {2007})},\ \Eprint {http://arxiv.org/abs/0705.2880}
  {arXiv:0705.2880 [gr-qc]} \BibitemShut {NoStop}%
\bibitem [{\citenamefont {Hartnoll}\ \emph {et~al.}(2008)\citenamefont
  {Hartnoll}, \citenamefont {Herzog},\ and\ \citenamefont
  {Horowitz}}]{Hartnoll:2008vx}%
  \BibitemOpen
  \bibfield  {author} {\bibinfo {author} {\bibfnamefont {Sean~A.}\ \bibnamefont
  {Hartnoll}}, \bibinfo {author} {\bibfnamefont {Christopher~P.}\ \bibnamefont
  {Herzog}}, \ and\ \bibinfo {author} {\bibfnamefont {Gary~T.}\ \bibnamefont
  {Horowitz}},\ }\bibfield  {title} {\enquote {\bibinfo {title} {{Building a
  Holographic Superconductor}},}\ }\href {\doibase
  10.1103/PhysRevLett.101.031601} {\bibfield  {journal} {\bibinfo  {journal}
  {Phys. Rev. Lett.}\ }\textbf {\bibinfo {volume} {101}},\ \bibinfo {pages}
  {031601} (\bibinfo {year} {2008})},\ \Eprint {http://arxiv.org/abs/0803.3295}
  {arXiv:0803.3295 [hep-th]} \BibitemShut {NoStop}%
\bibitem [{\citenamefont {Casals}\ \emph {et~al.}(2008)\citenamefont {Casals},
  \citenamefont {Dolan}, \citenamefont {Kanti},\ and\ \citenamefont
  {Winstanley}}]{Casals:2008pq}%
  \BibitemOpen
  \bibfield  {author} {\bibinfo {author} {\bibfnamefont {M.}~\bibnamefont
  {Casals}}, \bibinfo {author} {\bibfnamefont {S.~R.}\ \bibnamefont {Dolan}},
  \bibinfo {author} {\bibfnamefont {P.}~\bibnamefont {Kanti}}, \ and\ \bibinfo
  {author} {\bibfnamefont {E.}~\bibnamefont {Winstanley}},\ }\bibfield  {title}
  {\enquote {\bibinfo {title} {{Bulk Emission of Scalars by a Rotating Black
  Hole}},}\ }\href {\doibase 10.1088/1126-6708/2008/06/071} {\bibfield
  {journal} {\bibinfo  {journal} {JHEP}\ }\textbf {\bibinfo {volume} {06}},\
  \bibinfo {pages} {071} (\bibinfo {year} {2008})},\ \Eprint
  {http://arxiv.org/abs/0801.4910} {arXiv:0801.4910 [hep-th]} \BibitemShut
  {NoStop}%
\bibitem [{\citenamefont {Arvanitaki}\ \emph {et~al.}(2010)\citenamefont
  {Arvanitaki}, \citenamefont {Dimopoulos}, \citenamefont {Dubovsky},
  \citenamefont {Kaloper},\ and\ \citenamefont
  {March-Russell}}]{Arvanitaki:2009fg}%
  \BibitemOpen
  \bibfield  {author} {\bibinfo {author} {\bibfnamefont {Asimina}\ \bibnamefont
  {Arvanitaki}}, \bibinfo {author} {\bibfnamefont {Savas}\ \bibnamefont
  {Dimopoulos}}, \bibinfo {author} {\bibfnamefont {Sergei}\ \bibnamefont
  {Dubovsky}}, \bibinfo {author} {\bibfnamefont {Nemanja}\ \bibnamefont
  {Kaloper}}, \ and\ \bibinfo {author} {\bibfnamefont {John}\ \bibnamefont
  {March-Russell}},\ }\bibfield  {title} {\enquote {\bibinfo {title} {{String
  Axiverse}},}\ }\href {\doibase 10.1103/PhysRevD.81.123530} {\bibfield
  {journal} {\bibinfo  {journal} {Phys. Rev. D}\ }\textbf {\bibinfo {volume}
  {81}},\ \bibinfo {pages} {123530} (\bibinfo {year} {2010})},\ \Eprint
  {http://arxiv.org/abs/0905.4720} {arXiv:0905.4720 [hep-th]} \BibitemShut
  {NoStop}%
\bibitem [{\citenamefont {Arvanitaki}\ and\ \citenamefont
  {Dubovsky}(2011)}]{Arvanitaki:2010sy}%
  \BibitemOpen
  \bibfield  {author} {\bibinfo {author} {\bibfnamefont {Asimina}\ \bibnamefont
  {Arvanitaki}}\ and\ \bibinfo {author} {\bibfnamefont {Sergei}\ \bibnamefont
  {Dubovsky}},\ }\bibfield  {title} {\enquote {\bibinfo {title} {{Exploring the
  String Axiverse with Precision Black Hole Physics}},}\ }\href {\doibase
  10.1103/PhysRevD.83.044026} {\bibfield  {journal} {\bibinfo  {journal} {Phys.
  Rev. D}\ }\textbf {\bibinfo {volume} {83}},\ \bibinfo {pages} {044026}
  (\bibinfo {year} {2011})},\ \Eprint {http://arxiv.org/abs/1004.3558}
  {arXiv:1004.3558 [hep-th]} \BibitemShut {NoStop}%
\bibitem [{\citenamefont {Pani}\ \emph {et~al.}(2012)\citenamefont {Pani},
  \citenamefont {Cardoso}, \citenamefont {Gualtieri}, \citenamefont {Berti},\
  and\ \citenamefont {Ishibashi}}]{Pani:2012vp}%
  \BibitemOpen
  \bibfield  {author} {\bibinfo {author} {\bibfnamefont {Paolo}\ \bibnamefont
  {Pani}}, \bibinfo {author} {\bibfnamefont {Vitor}\ \bibnamefont {Cardoso}},
  \bibinfo {author} {\bibfnamefont {Leonardo}\ \bibnamefont {Gualtieri}},
  \bibinfo {author} {\bibfnamefont {Emanuele}\ \bibnamefont {Berti}}, \ and\
  \bibinfo {author} {\bibfnamefont {Akihiro}\ \bibnamefont {Ishibashi}},\
  }\bibfield  {title} {\enquote {\bibinfo {title} {{Black hole bombs and photon
  mass bounds}},}\ }\href {\doibase 10.1103/PhysRevLett.109.131102} {\bibfield
  {journal} {\bibinfo  {journal} {Phys. Rev. Lett.}\ }\textbf {\bibinfo
  {volume} {109}},\ \bibinfo {pages} {131102} (\bibinfo {year} {2012})},\
  \Eprint {http://arxiv.org/abs/1209.0465} {arXiv:1209.0465 [gr-qc]}
  \BibitemShut {NoStop}%
\bibitem [{\citenamefont {Witek}\ \emph {et~al.}(2013)\citenamefont {Witek},
  \citenamefont {Cardoso}, \citenamefont {Ishibashi},\ and\ \citenamefont
  {Sperhake}}]{Witek:2012tr}%
  \BibitemOpen
  \bibfield  {author} {\bibinfo {author} {\bibfnamefont {Helvi}\ \bibnamefont
  {Witek}}, \bibinfo {author} {\bibfnamefont {Vitor}\ \bibnamefont {Cardoso}},
  \bibinfo {author} {\bibfnamefont {Akihiro}\ \bibnamefont {Ishibashi}}, \ and\
  \bibinfo {author} {\bibfnamefont {Ulrich}\ \bibnamefont {Sperhake}},\
  }\bibfield  {title} {\enquote {\bibinfo {title} {{Superradiant instabilities
  in astrophysical systems}},}\ }\href {\doibase 10.1103/PhysRevD.87.043513}
  {\bibfield  {journal} {\bibinfo  {journal} {Phys. Rev. D}\ }\textbf {\bibinfo
  {volume} {87}},\ \bibinfo {pages} {043513} (\bibinfo {year} {2013})},\
  \Eprint {http://arxiv.org/abs/1212.0551} {arXiv:1212.0551 [gr-qc]}
  \BibitemShut {NoStop}%
\bibitem [{\citenamefont {Herdeiro}\ \emph {et~al.}(2013)\citenamefont
  {Herdeiro}, \citenamefont {Degollado},\ and\ \citenamefont
  {R\'unarsson}}]{Herdeiro:2013pia}%
  \BibitemOpen
  \bibfield  {author} {\bibinfo {author} {\bibfnamefont {Carlos A.~R.}\
  \bibnamefont {Herdeiro}}, \bibinfo {author} {\bibfnamefont {Juan~Carlos}\
  \bibnamefont {Degollado}}, \ and\ \bibinfo {author} {\bibfnamefont
  {Helgi~Freyr}\ \bibnamefont {R\'unarsson}},\ }\bibfield  {title} {\enquote
  {\bibinfo {title} {{Rapid growth of superradiant instabilities for charged
  black holes in a cavity}},}\ }\href {\doibase 10.1103/PhysRevD.88.063003}
  {\bibfield  {journal} {\bibinfo  {journal} {Phys. Rev. D}\ }\textbf {\bibinfo
  {volume} {88}},\ \bibinfo {pages} {063003} (\bibinfo {year} {2013})},\
  \Eprint {http://arxiv.org/abs/1305.5513} {arXiv:1305.5513 [gr-qc]}
  \BibitemShut {NoStop}%
\bibitem [{\citenamefont {Yoshino}\ and\ \citenamefont
  {Kodama}(2014)}]{Yoshino:2013ofa}%
  \BibitemOpen
  \bibfield  {author} {\bibinfo {author} {\bibfnamefont {Hirotaka}\
  \bibnamefont {Yoshino}}\ and\ \bibinfo {author} {\bibfnamefont {Hideo}\
  \bibnamefont {Kodama}},\ }\bibfield  {title} {\enquote {\bibinfo {title}
  {{Gravitational radiation from an axion cloud around a black hole:
  Superradiant phase}},}\ }\href {\doibase 10.1093/ptep/ptu029} {\bibfield
  {journal} {\bibinfo  {journal} {PTEP}\ }\textbf {\bibinfo {volume} {2014}},\
  \bibinfo {pages} {043E02} (\bibinfo {year} {2014})},\ \Eprint
  {http://arxiv.org/abs/1312.2326} {arXiv:1312.2326 [gr-qc]} \BibitemShut
  {NoStop}%
\bibitem [{\citenamefont {Basak}\ and\ \citenamefont
  {Majumdar}(2003)}]{Basak:2002aw}%
  \BibitemOpen
  \bibfield  {author} {\bibinfo {author} {\bibfnamefont {Soumen}\ \bibnamefont
  {Basak}}\ and\ \bibinfo {author} {\bibfnamefont {Parthasarathi}\ \bibnamefont
  {Majumdar}},\ }\bibfield  {title} {\enquote {\bibinfo {title}
  {{`Superresonance' from a rotating acoustic black hole}},}\ }\href {\doibase
  10.1088/0264-9381/20/18/304} {\bibfield  {journal} {\bibinfo  {journal}
  {Class. Quant. Grav.}\ }\textbf {\bibinfo {volume} {20}},\ \bibinfo {pages}
  {3907--3914} (\bibinfo {year} {2003})},\ \Eprint
  {http://arxiv.org/abs/gr-qc/0203059} {arXiv:gr-qc/0203059} \BibitemShut
  {NoStop}%
\bibitem [{\citenamefont {Richartz}\ \emph {et~al.}(2015)\citenamefont
  {Richartz}, \citenamefont {Prain}, \citenamefont {Liberati},\ and\
  \citenamefont {Weinfurtner}}]{Richartz:2014lda}%
  \BibitemOpen
  \bibfield  {author} {\bibinfo {author} {\bibfnamefont {Maur\'\i{}cio}\
  \bibnamefont {Richartz}}, \bibinfo {author} {\bibfnamefont {Angus}\
  \bibnamefont {Prain}}, \bibinfo {author} {\bibfnamefont {Stefano}\
  \bibnamefont {Liberati}}, \ and\ \bibinfo {author} {\bibfnamefont {Silke}\
  \bibnamefont {Weinfurtner}},\ }\bibfield  {title} {\enquote {\bibinfo {title}
  {{Rotating black holes in a draining bathtub: superradiant scattering of
  gravity waves}},}\ }\href {\doibase 10.1103/PhysRevD.91.124018} {\bibfield
  {journal} {\bibinfo  {journal} {Phys. Rev. D}\ }\textbf {\bibinfo {volume}
  {91}},\ \bibinfo {pages} {124018} (\bibinfo {year} {2015})},\ \Eprint
  {http://arxiv.org/abs/1411.1662} {arXiv:1411.1662 [gr-qc]} \BibitemShut
  {NoStop}%
\bibitem [{\citenamefont {Torres}\ \emph {et~al.}(2017)\citenamefont {Torres},
  \citenamefont {Patrick}, \citenamefont {Coutant}, \citenamefont {Richartz},
  \citenamefont {Tedford},\ and\ \citenamefont {Weinfurtner}}]{Torres:2016iee}%
  \BibitemOpen
  \bibfield  {author} {\bibinfo {author} {\bibfnamefont {Theo}\ \bibnamefont
  {Torres}}, \bibinfo {author} {\bibfnamefont {Sam}\ \bibnamefont {Patrick}},
  \bibinfo {author} {\bibfnamefont {Antonin}\ \bibnamefont {Coutant}}, \bibinfo
  {author} {\bibfnamefont {Mauricio}\ \bibnamefont {Richartz}}, \bibinfo
  {author} {\bibfnamefont {Edmund~W.}\ \bibnamefont {Tedford}}, \ and\ \bibinfo
  {author} {\bibfnamefont {Silke}\ \bibnamefont {Weinfurtner}},\ }\bibfield
  {title} {\enquote {\bibinfo {title} {{Observation of superradiance in a
  vortex flow}},}\ }\href {\doibase 10.1038/nphys4151} {\bibfield  {journal}
  {\bibinfo  {journal} {Nature Phys.}\ }\textbf {\bibinfo {volume} {13}},\
  \bibinfo {pages} {833--836} (\bibinfo {year} {2017})},\ \Eprint
  {http://arxiv.org/abs/1612.06180} {arXiv:1612.06180 [gr-qc]} \BibitemShut
  {NoStop}%
\bibitem [{\citenamefont {Benone}\ and\ \citenamefont
  {Crispino}(2016)}]{Benone:2015bst}%
  \BibitemOpen
  \bibfield  {author} {\bibinfo {author} {\bibfnamefont {Carolina~L.}\
  \bibnamefont {Benone}}\ and\ \bibinfo {author} {\bibfnamefont {Lu\'\i{}s
  C.~B.}\ \bibnamefont {Crispino}},\ }\bibfield  {title} {\enquote {\bibinfo
  {title} {{Superradiance in static black hole spacetimes}},}\ }\href {\doibase
  10.1103/PhysRevD.93.024028} {\bibfield  {journal} {\bibinfo  {journal} {Phys.
  Rev. D}\ }\textbf {\bibinfo {volume} {93}},\ \bibinfo {pages} {024028}
  (\bibinfo {year} {2016})},\ \Eprint {http://arxiv.org/abs/1511.02634}
  {arXiv:1511.02634 [gr-qc]} \BibitemShut {NoStop}%
\bibitem [{\citenamefont {Konoplya}\ and\ \citenamefont
  {Zhidenko}(2016)}]{Konoplya:2016hmd}%
  \BibitemOpen
  \bibfield  {author} {\bibinfo {author} {\bibfnamefont {R.~A.}\ \bibnamefont
  {Konoplya}}\ and\ \bibinfo {author} {\bibfnamefont {A.}~\bibnamefont
  {Zhidenko}},\ }\bibfield  {title} {\enquote {\bibinfo {title} {{Wormholes
  versus black holes: quasinormal ringing at early and late times}},}\ }\href
  {\doibase 10.1088/1475-7516/2016/12/043} {\bibfield  {journal} {\bibinfo
  {journal} {JCAP}\ }\textbf {\bibinfo {volume} {12}},\ \bibinfo {pages} {043}
  (\bibinfo {year} {2016})},\ \Eprint {http://arxiv.org/abs/1606.00517}
  {arXiv:1606.00517 [gr-qc]} \BibitemShut {NoStop}%
\bibitem [{\citenamefont {Hod}(2016)}]{Hod:2016iri}%
  \BibitemOpen
  \bibfield  {author} {\bibinfo {author} {\bibfnamefont {Shahar}\ \bibnamefont
  {Hod}},\ }\bibfield  {title} {\enquote {\bibinfo {title} {{The superradiant
  instability regime of the spinning Kerr black hole}},}\ }\href {\doibase
  10.1016/j.physletb.2016.05.012} {\bibfield  {journal} {\bibinfo  {journal}
  {Phys. Lett. B}\ }\textbf {\bibinfo {volume} {758}},\ \bibinfo {pages}
  {181--185} (\bibinfo {year} {2016})},\ \Eprint
  {http://arxiv.org/abs/1606.02306} {arXiv:1606.02306 [gr-qc]} \BibitemShut
  {NoStop}%
\bibitem [{\citenamefont {Baryakhtar}\ \emph {et~al.}(2017)\citenamefont
  {Baryakhtar}, \citenamefont {Lasenby},\ and\ \citenamefont
  {Teo}}]{Baryakhtar:2017ngi}%
  \BibitemOpen
  \bibfield  {author} {\bibinfo {author} {\bibfnamefont {Masha}\ \bibnamefont
  {Baryakhtar}}, \bibinfo {author} {\bibfnamefont {Robert}\ \bibnamefont
  {Lasenby}}, \ and\ \bibinfo {author} {\bibfnamefont {Mae}\ \bibnamefont
  {Teo}},\ }\bibfield  {title} {\enquote {\bibinfo {title} {{Black Hole
  Superradiance Signatures of Ultralight Vectors}},}\ }\href {\doibase
  10.1103/PhysRevD.96.035019} {\bibfield  {journal} {\bibinfo  {journal} {Phys.
  Rev. D}\ }\textbf {\bibinfo {volume} {96}},\ \bibinfo {pages} {035019}
  (\bibinfo {year} {2017})},\ \Eprint {http://arxiv.org/abs/1704.05081}
  {arXiv:1704.05081 [hep-ph]} \BibitemShut {NoStop}%
\bibitem [{\citenamefont {East}\ and\ \citenamefont
  {Pretorius}(2017)}]{East:2017ovw}%
  \BibitemOpen
  \bibfield  {author} {\bibinfo {author} {\bibfnamefont {William~E.}\
  \bibnamefont {East}}\ and\ \bibinfo {author} {\bibfnamefont {Frans}\
  \bibnamefont {Pretorius}},\ }\bibfield  {title} {\enquote {\bibinfo {title}
  {{Superradiant Instability and Backreaction of Massive Vector Fields around
  Kerr Black Holes}},}\ }\href {\doibase 10.1103/PhysRevLett.119.041101}
  {\bibfield  {journal} {\bibinfo  {journal} {Phys. Rev. Lett.}\ }\textbf
  {\bibinfo {volume} {119}},\ \bibinfo {pages} {041101} (\bibinfo {year}
  {2017})},\ \Eprint {http://arxiv.org/abs/1704.04791} {arXiv:1704.04791
  [gr-qc]} \BibitemShut {NoStop}%
\bibitem [{\citenamefont {Rosa}\ and\ \citenamefont
  {Kephart}(2018)}]{Rosa:2017ury}%
  \BibitemOpen
  \bibfield  {author} {\bibinfo {author} {\bibfnamefont {Jo\~ao~G.}\
  \bibnamefont {Rosa}}\ and\ \bibinfo {author} {\bibfnamefont {Thomas~W.}\
  \bibnamefont {Kephart}},\ }\bibfield  {title} {\enquote {\bibinfo {title}
  {{Stimulated Axion Decay in Superradiant Clouds around Primordial Black
  Holes}},}\ }\href {\doibase 10.1103/PhysRevLett.120.231102} {\bibfield
  {journal} {\bibinfo  {journal} {Phys. Rev. Lett.}\ }\textbf {\bibinfo
  {volume} {120}},\ \bibinfo {pages} {231102} (\bibinfo {year} {2018})},\
  \Eprint {http://arxiv.org/abs/1709.06581} {arXiv:1709.06581 [gr-qc]}
  \BibitemShut {NoStop}%
\bibitem [{\citenamefont {Cardoso}\ \emph {et~al.}(2018)\citenamefont
  {Cardoso}, \citenamefont {Dias}, \citenamefont {Hartnett}, \citenamefont
  {Middleton}, \citenamefont {Pani},\ and\ \citenamefont
  {Santos}}]{Cardoso:2018tly}%
  \BibitemOpen
  \bibfield  {author} {\bibinfo {author} {\bibfnamefont {Vitor}\ \bibnamefont
  {Cardoso}}, \bibinfo {author} {\bibfnamefont {\'Oscar J.~C.}\ \bibnamefont
  {Dias}}, \bibinfo {author} {\bibfnamefont {Gavin~S.}\ \bibnamefont
  {Hartnett}}, \bibinfo {author} {\bibfnamefont {Matthew}\ \bibnamefont
  {Middleton}}, \bibinfo {author} {\bibfnamefont {Paolo}\ \bibnamefont {Pani}},
  \ and\ \bibinfo {author} {\bibfnamefont {Jorge~E.}\ \bibnamefont {Santos}},\
  }\bibfield  {title} {\enquote {\bibinfo {title} {{Constraining the mass of
  dark photons and axion-like particles through black-hole superradiance}},}\
  }\href {\doibase 10.1088/1475-7516/2018/03/043} {\bibfield  {journal}
  {\bibinfo  {journal} {JCAP}\ }\textbf {\bibinfo {volume} {03}},\ \bibinfo
  {pages} {043} (\bibinfo {year} {2018})},\ \Eprint
  {http://arxiv.org/abs/1801.01420} {arXiv:1801.01420 [gr-qc]} \BibitemShut
  {NoStop}%
\bibitem [{\citenamefont {Degollado}\ \emph {et~al.}(2018)\citenamefont
  {Degollado}, \citenamefont {Herdeiro},\ and\ \citenamefont
  {Radu}}]{Degollado:2018ypf}%
  \BibitemOpen
  \bibfield  {author} {\bibinfo {author} {\bibfnamefont {Juan~Carlos}\
  \bibnamefont {Degollado}}, \bibinfo {author} {\bibfnamefont {Carlos A.~R.}\
  \bibnamefont {Herdeiro}}, \ and\ \bibinfo {author} {\bibfnamefont {Eugen}\
  \bibnamefont {Radu}},\ }\bibfield  {title} {\enquote {\bibinfo {title}
  {{Effective stability against superradiance of Kerr black holes with
  synchronised hair}},}\ }\href {\doibase 10.1016/j.physletb.2018.04.052}
  {\bibfield  {journal} {\bibinfo  {journal} {Phys. Lett. B}\ }\textbf
  {\bibinfo {volume} {781}},\ \bibinfo {pages} {651--655} (\bibinfo {year}
  {2018})},\ \Eprint {http://arxiv.org/abs/1802.07266} {arXiv:1802.07266
  [gr-qc]} \BibitemShut {NoStop}%
\bibitem [{\citenamefont {Zhang}\ \emph {et~al.}(2020)\citenamefont {Zhang},
  \citenamefont {Zhang}, \citenamefont {Li},\ and\ \citenamefont
  {Guo}}]{Zhang:2020sjh}%
  \BibitemOpen
  \bibfield  {author} {\bibinfo {author} {\bibfnamefont {Cheng-Yong}\
  \bibnamefont {Zhang}}, \bibinfo {author} {\bibfnamefont {Shao-Jun}\
  \bibnamefont {Zhang}}, \bibinfo {author} {\bibfnamefont {Peng-Cheng}\
  \bibnamefont {Li}}, \ and\ \bibinfo {author} {\bibfnamefont {Minyong}\
  \bibnamefont {Guo}},\ }\bibfield  {title} {\enquote {\bibinfo {title}
  {{Superradiance and stability of the regularized 4D charged
  Einstein-Gauss-Bonnet black hole}},}\ }\href {\doibase
  10.1007/JHEP08(2020)105} {\bibfield  {journal} {\bibinfo  {journal} {JHEP}\
  }\textbf {\bibinfo {volume} {08}},\ \bibinfo {pages} {105} (\bibinfo {year}
  {2020})},\ \Eprint {http://arxiv.org/abs/2004.03141} {arXiv:2004.03141
  [gr-qc]} \BibitemShut {NoStop}%
\bibitem [{\citenamefont {Mehta}\ \emph {et~al.}(2021)\citenamefont {Mehta},
  \citenamefont {Demirtas}, \citenamefont {Long}, \citenamefont {Marsh},
  \citenamefont {McAllister},\ and\ \citenamefont {Stott}}]{Mehta:2021pwf}%
  \BibitemOpen
  \bibfield  {author} {\bibinfo {author} {\bibfnamefont {Viraf~M.}\
  \bibnamefont {Mehta}}, \bibinfo {author} {\bibfnamefont {Mehmet}\
  \bibnamefont {Demirtas}}, \bibinfo {author} {\bibfnamefont {Cody}\
  \bibnamefont {Long}}, \bibinfo {author} {\bibfnamefont {David J.~E.}\
  \bibnamefont {Marsh}}, \bibinfo {author} {\bibfnamefont {Liam}\ \bibnamefont
  {McAllister}}, \ and\ \bibinfo {author} {\bibfnamefont {Matthew~J.}\
  \bibnamefont {Stott}},\ }\bibfield  {title} {\enquote {\bibinfo {title}
  {{Superradiance in string theory}},}\ }\href {\doibase
  10.1088/1475-7516/2021/07/033} {\bibfield  {journal} {\bibinfo  {journal}
  {JCAP}\ }\textbf {\bibinfo {volume} {07}},\ \bibinfo {pages} {033} (\bibinfo
  {year} {2021})},\ \Eprint {http://arxiv.org/abs/2103.06812} {arXiv:2103.06812
  [hep-th]} \BibitemShut {NoStop}%
\bibitem [{\citenamefont {Richartz}\ and\ \citenamefont
  {Saa}(2013)}]{Richartz:2013unq}%
  \BibitemOpen
  \bibfield  {author} {\bibinfo {author} {\bibfnamefont {Maur\'\i{}cio}\
  \bibnamefont {Richartz}}\ and\ \bibinfo {author} {\bibfnamefont {Alberto}\
  \bibnamefont {Saa}},\ }\bibfield  {title} {\enquote {\bibinfo {title}
  {{Superradiance without event horizons in General Relativity}},}\ }\href
  {\doibase 10.1103/PhysRevD.88.044008} {\bibfield  {journal} {\bibinfo
  {journal} {Phys. Rev. D}\ }\textbf {\bibinfo {volume} {88}},\ \bibinfo
  {pages} {044008} (\bibinfo {year} {2013})},\ \Eprint
  {http://arxiv.org/abs/1306.3137} {arXiv:1306.3137 [gr-qc]} \BibitemShut
  {NoStop}%
\bibitem [{\citenamefont {Cardoso}\ \emph {et~al.}(2015)\citenamefont
  {Cardoso}, \citenamefont {Brito},\ and\ \citenamefont
  {Rosa}}]{Cardoso:2015zqa}%
  \BibitemOpen
  \bibfield  {author} {\bibinfo {author} {\bibfnamefont {Vitor}\ \bibnamefont
  {Cardoso}}, \bibinfo {author} {\bibfnamefont {Richard}\ \bibnamefont
  {Brito}}, \ and\ \bibinfo {author} {\bibfnamefont {Joao~L.}\ \bibnamefont
  {Rosa}},\ }\bibfield  {title} {\enquote {\bibinfo {title} {{Superradiance in
  stars}},}\ }\href {\doibase 10.1103/PhysRevD.91.124026} {\bibfield  {journal}
  {\bibinfo  {journal} {Phys. Rev. D}\ }\textbf {\bibinfo {volume} {91}},\
  \bibinfo {pages} {124026} (\bibinfo {year} {2015})},\ \Eprint
  {http://arxiv.org/abs/1505.05509} {arXiv:1505.05509 [gr-qc]} \BibitemShut
  {NoStop}%
\bibitem [{\citenamefont {Cardoso}\ \emph {et~al.}(2017)\citenamefont
  {Cardoso}, \citenamefont {Pani},\ and\ \citenamefont {Yu}}]{Cardoso:2017kgn}%
  \BibitemOpen
  \bibfield  {author} {\bibinfo {author} {\bibfnamefont {Vitor}\ \bibnamefont
  {Cardoso}}, \bibinfo {author} {\bibfnamefont {Paolo}\ \bibnamefont {Pani}}, \
  and\ \bibinfo {author} {\bibfnamefont {Tien-Tien}\ \bibnamefont {Yu}},\
  }\bibfield  {title} {\enquote {\bibinfo {title} {{Superradiance in rotating
  stars and pulsar-timing constraints on dark photons}},}\ }\href {\doibase
  10.1103/PhysRevD.95.124056} {\bibfield  {journal} {\bibinfo  {journal} {Phys.
  Rev. D}\ }\textbf {\bibinfo {volume} {95}},\ \bibinfo {pages} {124056}
  (\bibinfo {year} {2017})},\ \Eprint {http://arxiv.org/abs/1704.06151}
  {arXiv:1704.06151 [gr-qc]} \BibitemShut {NoStop}%
\bibitem [{\citenamefont {Day}\ and\ \citenamefont
  {McDonald}(2019)}]{Day:2019bbh}%
  \BibitemOpen
  \bibfield  {author} {\bibinfo {author} {\bibfnamefont {Francesca~V.}\
  \bibnamefont {Day}}\ and\ \bibinfo {author} {\bibfnamefont {Jamie~I.}\
  \bibnamefont {McDonald}},\ }\bibfield  {title} {\enquote {\bibinfo {title}
  {{Axion superradiance in rotating neutron stars}},}\ }\href {\doibase
  10.1088/1475-7516/2019/10/051} {\bibfield  {journal} {\bibinfo  {journal}
  {JCAP}\ }\textbf {\bibinfo {volume} {10}},\ \bibinfo {pages} {051} (\bibinfo
  {year} {2019})},\ \Eprint {http://arxiv.org/abs/1904.08341} {arXiv:1904.08341
  [hep-ph]} \BibitemShut {NoStop}%
\bibitem [{\citenamefont {Chadha-Day}\ \emph {et~al.}(2022)\citenamefont
  {Chadha-Day}, \citenamefont {Garbrecht},\ and\ \citenamefont
  {McDonald}}]{Chadha-Day:2022inf}%
  \BibitemOpen
  \bibfield  {author} {\bibinfo {author} {\bibfnamefont {Francesca}\
  \bibnamefont {Chadha-Day}}, \bibinfo {author} {\bibfnamefont {Bj\"orn}\
  \bibnamefont {Garbrecht}}, \ and\ \bibinfo {author} {\bibfnamefont {Jamie}\
  \bibnamefont {McDonald}},\ }\bibfield  {title} {\enquote {\bibinfo {title}
  {{Superradiance in stars: non-equilibrium approach to damping of fields in
  stellar media}},}\ }\href {\doibase 10.1088/1475-7516/2022/12/008} {\bibfield
   {journal} {\bibinfo  {journal} {JCAP}\ }\textbf {\bibinfo {volume} {12}},\
  \bibinfo {pages} {008} (\bibinfo {year} {2022})},\ \Eprint
  {http://arxiv.org/abs/2207.07662} {arXiv:2207.07662 [hep-ph]} \BibitemShut
  {NoStop}%
\bibitem [{\citenamefont {Bekenstein}\ and\ \citenamefont
  {Schiffer}(1998)}]{Bekenstein:1998nt}%
  \BibitemOpen
  \bibfield  {author} {\bibinfo {author} {\bibfnamefont {Jacob~D.}\
  \bibnamefont {Bekenstein}}\ and\ \bibinfo {author} {\bibfnamefont {Marcelo}\
  \bibnamefont {Schiffer}},\ }\bibfield  {title} {\enquote {\bibinfo {title}
  {{The Many faces of superradiance}},}\ }\href {\doibase
  10.1103/PhysRevD.58.064014} {\bibfield  {journal} {\bibinfo  {journal} {Phys.
  Rev. D}\ }\textbf {\bibinfo {volume} {58}},\ \bibinfo {pages} {064014}
  (\bibinfo {year} {1998})},\ \Eprint {http://arxiv.org/abs/gr-qc/9803033}
  {arXiv:gr-qc/9803033} \BibitemShut {NoStop}%
\bibitem [{\citenamefont {Brito}\ \emph {et~al.}(2015)\citenamefont {Brito},
  \citenamefont {Cardoso},\ and\ \citenamefont {Pani}}]{Brito:2015oca}%
  \BibitemOpen
  \bibfield  {author} {\bibinfo {author} {\bibfnamefont {Richard}\ \bibnamefont
  {Brito}}, \bibinfo {author} {\bibfnamefont {Vitor}\ \bibnamefont {Cardoso}},
  \ and\ \bibinfo {author} {\bibfnamefont {Paolo}\ \bibnamefont {Pani}},\
  }\bibfield  {title} {\enquote {\bibinfo {title} {{Superradiance}: {New
  Frontiers in Black Hole Physics}},}\ }\href {\doibase
  10.1007/978-3-319-19000-6} {\bibfield  {journal} {\bibinfo  {journal} {Lect.
  Notes Phys.}\ }\textbf {\bibinfo {volume} {906}},\ \bibinfo {pages}
  {pp.1--237} (\bibinfo {year} {2015})},\ \Eprint
  {http://arxiv.org/abs/1501.06570} {arXiv:1501.06570 [gr-qc]} \BibitemShut
  {NoStop}%
\bibitem [{\citenamefont {Rosen}(1968)}]{Rosen:1968mfz}%
  \BibitemOpen
  \bibfield  {author} {\bibinfo {author} {\bibfnamefont {Gerald}\ \bibnamefont
  {Rosen}},\ }\bibfield  {title} {\enquote {\bibinfo {title} {{Particlelike
  Solutions to Nonlinear Complex Scalar Field Theories with Positive-Definite
  Energy Densities}},}\ }\href {\doibase 10.1063/1.1664693} {\bibfield
  {journal} {\bibinfo  {journal} {J. Math. Phys.}\ }\textbf {\bibinfo {volume}
  {9}},\ \bibinfo {pages} {996} (\bibinfo {year} {1968})}\BibitemShut {NoStop}%
\bibitem [{\citenamefont {Friedberg}\ \emph {et~al.}(1976)\citenamefont
  {Friedberg}, \citenamefont {Lee},\ and\ \citenamefont
  {Sirlin}}]{Friedberg:1976me}%
  \BibitemOpen
  \bibfield  {author} {\bibinfo {author} {\bibfnamefont {R.}~\bibnamefont
  {Friedberg}}, \bibinfo {author} {\bibfnamefont {T.~D.}\ \bibnamefont {Lee}},
  \ and\ \bibinfo {author} {\bibfnamefont {A.}~\bibnamefont {Sirlin}},\
  }\bibfield  {title} {\enquote {\bibinfo {title} {{A Class of Scalar-Field
  Soliton Solutions in Three Space Dimensions}},}\ }\href {\doibase
  10.1103/PhysRevD.13.2739} {\bibfield  {journal} {\bibinfo  {journal} {Phys.
  Rev. D}\ }\textbf {\bibinfo {volume} {13}},\ \bibinfo {pages} {2739--2761}
  (\bibinfo {year} {1976})}\BibitemShut {NoStop}%
\bibitem [{\citenamefont {Coleman}(1985)}]{Coleman:1985ki}%
  \BibitemOpen
  \bibfield  {author} {\bibinfo {author} {\bibfnamefont {Sidney~R.}\
  \bibnamefont {Coleman}},\ }\bibfield  {title} {\enquote {\bibinfo {title}
  {{Q-balls}},}\ }\href {\doibase 10.1016/0550-3213(86)90520-1} {\bibfield
  {journal} {\bibinfo  {journal} {Nucl. Phys. B}\ }\textbf {\bibinfo {volume}
  {262}},\ \bibinfo {pages} {263} (\bibinfo {year} {1985})},\ \bibinfo {note}
  {[Addendum: Nucl.Phys.B 269, 744 (1986)]}\BibitemShut {NoStop}%
\bibitem [{\citenamefont {Lee}\ and\ \citenamefont {Pang}(1992)}]{Lee:1991ax}%
  \BibitemOpen
  \bibfield  {author} {\bibinfo {author} {\bibfnamefont {T.~D.}\ \bibnamefont
  {Lee}}\ and\ \bibinfo {author} {\bibfnamefont {Y.}~\bibnamefont {Pang}},\
  }\bibfield  {title} {\enquote {\bibinfo {title} {{Nontopological
  solitons}},}\ }\href {\doibase 10.1016/0370-1573(92)90064-7} {\bibfield
  {journal} {\bibinfo  {journal} {Phys. Rept.}\ }\textbf {\bibinfo {volume}
  {221}},\ \bibinfo {pages} {251--350} (\bibinfo {year} {1992})}\BibitemShut
  {NoStop}%
\bibitem [{\citenamefont {Enqvist}\ and\ \citenamefont
  {McDonald}(1998)}]{Enqvist:1997si}%
  \BibitemOpen
  \bibfield  {author} {\bibinfo {author} {\bibfnamefont {Kari}\ \bibnamefont
  {Enqvist}}\ and\ \bibinfo {author} {\bibfnamefont {John}\ \bibnamefont
  {McDonald}},\ }\bibfield  {title} {\enquote {\bibinfo {title} {{Q balls and
  baryogenesis in the MSSM}},}\ }\href {\doibase 10.1016/S0370-2693(98)00271-8}
  {\bibfield  {journal} {\bibinfo  {journal} {Phys. Lett. B}\ }\textbf
  {\bibinfo {volume} {425}},\ \bibinfo {pages} {309--321} (\bibinfo {year}
  {1998})},\ \Eprint {http://arxiv.org/abs/hep-ph/9711514}
  {arXiv:hep-ph/9711514} \BibitemShut {NoStop}%
\bibitem [{\citenamefont {Kasuya}\ and\ \citenamefont
  {Kawasaki}(2000)}]{Kasuya:1999wu}%
  \BibitemOpen
  \bibfield  {author} {\bibinfo {author} {\bibfnamefont {S.}~\bibnamefont
  {Kasuya}}\ and\ \bibinfo {author} {\bibfnamefont {M.}~\bibnamefont
  {Kawasaki}},\ }\bibfield  {title} {\enquote {\bibinfo {title} {{Q ball
  formation through Affleck-Dine mechanism}},}\ }\href {\doibase
  10.1103/PhysRevD.61.041301} {\bibfield  {journal} {\bibinfo  {journal} {Phys.
  Rev. D}\ }\textbf {\bibinfo {volume} {61}},\ \bibinfo {pages} {041301}
  (\bibinfo {year} {2000})},\ \Eprint {http://arxiv.org/abs/hep-ph/9909509}
  {arXiv:hep-ph/9909509} \BibitemShut {NoStop}%
\bibitem [{\citenamefont {Multamaki}\ and\ \citenamefont
  {Vilja}(2002)}]{Multamaki:2002hv}%
  \BibitemOpen
  \bibfield  {author} {\bibinfo {author} {\bibfnamefont {Tuomas}\ \bibnamefont
  {Multamaki}}\ and\ \bibinfo {author} {\bibfnamefont {Iiro}\ \bibnamefont
  {Vilja}},\ }\bibfield  {title} {\enquote {\bibinfo {title} {{Simulations of Q
  ball formation}},}\ }\href {\doibase 10.1016/S0370-2693(02)01730-6}
  {\bibfield  {journal} {\bibinfo  {journal} {Phys. Lett. B}\ }\textbf
  {\bibinfo {volume} {535}},\ \bibinfo {pages} {170--176} (\bibinfo {year}
  {2002})},\ \Eprint {http://arxiv.org/abs/hep-ph/0203195}
  {arXiv:hep-ph/0203195} \BibitemShut {NoStop}%
\bibitem [{\citenamefont {Harigaya}\ \emph {et~al.}(2014)\citenamefont
  {Harigaya}, \citenamefont {Kamada}, \citenamefont {Kawasaki}, \citenamefont
  {Mukaida},\ and\ \citenamefont {Yamada}}]{Harigaya:2014tla}%
  \BibitemOpen
  \bibfield  {author} {\bibinfo {author} {\bibfnamefont {Keisuke}\ \bibnamefont
  {Harigaya}}, \bibinfo {author} {\bibfnamefont {Ayuki}\ \bibnamefont
  {Kamada}}, \bibinfo {author} {\bibfnamefont {Masahiro}\ \bibnamefont
  {Kawasaki}}, \bibinfo {author} {\bibfnamefont {Kyohei}\ \bibnamefont
  {Mukaida}}, \ and\ \bibinfo {author} {\bibfnamefont {Masaki}\ \bibnamefont
  {Yamada}},\ }\bibfield  {title} {\enquote {\bibinfo {title} {{Affleck-Dine
  Baryogenesis and Dark Matter Production after High-scale Inflation}},}\
  }\href {\doibase 10.1103/PhysRevD.90.043510} {\bibfield  {journal} {\bibinfo
  {journal} {Phys. Rev. D}\ }\textbf {\bibinfo {volume} {90}},\ \bibinfo
  {pages} {043510} (\bibinfo {year} {2014})},\ \Eprint
  {http://arxiv.org/abs/1404.3138} {arXiv:1404.3138 [hep-ph]} \BibitemShut
  {NoStop}%
\bibitem [{\citenamefont {Zhou}(2015)}]{Zhou:2015yfa}%
  \BibitemOpen
  \bibfield  {author} {\bibinfo {author} {\bibfnamefont {Shuang-Yong}\
  \bibnamefont {Zhou}},\ }\bibfield  {title} {\enquote {\bibinfo {title}
  {{Gravitational waves from Affleck-Dine condensate fragmentation}},}\ }\href
  {\doibase 10.1088/1475-7516/2015/06/033} {\bibfield  {journal} {\bibinfo
  {journal} {JCAP}\ }\textbf {\bibinfo {volume} {06}},\ \bibinfo {pages} {033}
  (\bibinfo {year} {2015})},\ \Eprint {http://arxiv.org/abs/1501.01217}
  {arXiv:1501.01217 [astro-ph.CO]} \BibitemShut {NoStop}%
\bibitem [{\citenamefont {Kusenko}\ and\ \citenamefont
  {Shaposhnikov}(1998)}]{Kusenko:1997si}%
  \BibitemOpen
  \bibfield  {author} {\bibinfo {author} {\bibfnamefont {Alexander}\
  \bibnamefont {Kusenko}}\ and\ \bibinfo {author} {\bibfnamefont {Mikhail~E.}\
  \bibnamefont {Shaposhnikov}},\ }\bibfield  {title} {\enquote {\bibinfo
  {title} {{Supersymmetric Q balls as dark matter}},}\ }\href {\doibase
  10.1016/S0370-2693(97)01375-0} {\bibfield  {journal} {\bibinfo  {journal}
  {Phys. Lett. B}\ }\textbf {\bibinfo {volume} {418}},\ \bibinfo {pages}
  {46--54} (\bibinfo {year} {1998})},\ \Eprint
  {http://arxiv.org/abs/hep-ph/9709492} {arXiv:hep-ph/9709492} \BibitemShut
  {NoStop}%
\bibitem [{\citenamefont {Banerjee}\ and\ \citenamefont
  {Jedamzik}(2000)}]{Banerjee:2000mb}%
  \BibitemOpen
  \bibfield  {author} {\bibinfo {author} {\bibfnamefont {Robi}\ \bibnamefont
  {Banerjee}}\ and\ \bibinfo {author} {\bibfnamefont {Karsten}\ \bibnamefont
  {Jedamzik}},\ }\bibfield  {title} {\enquote {\bibinfo {title} {{On B-ball
  dark matter and baryogenesis}},}\ }\href {\doibase
  10.1016/S0370-2693(00)00688-2} {\bibfield  {journal} {\bibinfo  {journal}
  {Phys. Lett. B}\ }\textbf {\bibinfo {volume} {484}},\ \bibinfo {pages}
  {278--282} (\bibinfo {year} {2000})},\ \Eprint
  {http://arxiv.org/abs/hep-ph/0005031} {arXiv:hep-ph/0005031} \BibitemShut
  {NoStop}%
\bibitem [{\citenamefont {Roszkowski}\ and\ \citenamefont
  {Seto}(2007)}]{Roszkowski:2006kw}%
  \BibitemOpen
  \bibfield  {author} {\bibinfo {author} {\bibfnamefont {Leszek}\ \bibnamefont
  {Roszkowski}}\ and\ \bibinfo {author} {\bibfnamefont {Osamu}\ \bibnamefont
  {Seto}},\ }\bibfield  {title} {\enquote {\bibinfo {title} {{Axino dark matter
  from Q-balls in Affleck-Dine baryogenesis and the Omega(b) - Omega(DM)
  coincidence problem}},}\ }\href {\doibase 10.1103/PhysRevLett.98.161304}
  {\bibfield  {journal} {\bibinfo  {journal} {Phys. Rev. Lett.}\ }\textbf
  {\bibinfo {volume} {98}},\ \bibinfo {pages} {161304} (\bibinfo {year}
  {2007})},\ \Eprint {http://arxiv.org/abs/hep-ph/0608013}
  {arXiv:hep-ph/0608013} \BibitemShut {NoStop}%
\bibitem [{\citenamefont {Shoemaker}\ and\ \citenamefont
  {Kusenko}(2009)}]{Shoemaker:2009kg}%
  \BibitemOpen
  \bibfield  {author} {\bibinfo {author} {\bibfnamefont {Ian~M.}\ \bibnamefont
  {Shoemaker}}\ and\ \bibinfo {author} {\bibfnamefont {Alexander}\ \bibnamefont
  {Kusenko}},\ }\bibfield  {title} {\enquote {\bibinfo {title} {{Gravitino dark
  matter from Q-ball decays}},}\ }\href {\doibase 10.1103/PhysRevD.80.075021}
  {\bibfield  {journal} {\bibinfo  {journal} {Phys. Rev. D}\ }\textbf {\bibinfo
  {volume} {80}},\ \bibinfo {pages} {075021} (\bibinfo {year} {2009})},\
  \Eprint {http://arxiv.org/abs/0909.3334} {arXiv:0909.3334 [hep-ph]}
  \BibitemShut {NoStop}%
\bibitem [{\citenamefont {Kasuya}\ \emph {et~al.}(2013)\citenamefont {Kasuya},
  \citenamefont {Kawasaki},\ and\ \citenamefont {Yamada}}]{Kasuya:2012mh}%
  \BibitemOpen
  \bibfield  {author} {\bibinfo {author} {\bibfnamefont {Shinta}\ \bibnamefont
  {Kasuya}}, \bibinfo {author} {\bibfnamefont {Masahiro}\ \bibnamefont
  {Kawasaki}}, \ and\ \bibinfo {author} {\bibfnamefont {Masaki}\ \bibnamefont
  {Yamada}},\ }\bibfield  {title} {\enquote {\bibinfo {title} {{Revisiting the
  gravitino dark matter and baryon asymmetry from Q-ball decay in gauge
  mediation}},}\ }\href {\doibase 10.1016/j.physletb.2013.08.008} {\bibfield
  {journal} {\bibinfo  {journal} {Phys. Lett. B}\ }\textbf {\bibinfo {volume}
  {726}},\ \bibinfo {pages} {1--7} (\bibinfo {year} {2013})},\ \Eprint
  {http://arxiv.org/abs/1211.4743} {arXiv:1211.4743 [hep-ph]} \BibitemShut
  {NoStop}%
\bibitem [{\citenamefont {Copeland}\ \emph {et~al.}(2014)\citenamefont
  {Copeland}, \citenamefont {Saffin},\ and\ \citenamefont
  {Zhou}}]{Copeland:2014qra}%
  \BibitemOpen
  \bibfield  {author} {\bibinfo {author} {\bibfnamefont {Edmund~J.}\
  \bibnamefont {Copeland}}, \bibinfo {author} {\bibfnamefont {Paul~M.}\
  \bibnamefont {Saffin}}, \ and\ \bibinfo {author} {\bibfnamefont
  {Shuang-Yong}\ \bibnamefont {Zhou}},\ }\bibfield  {title} {\enquote {\bibinfo
  {title} {{Charge-Swapping Q-balls}},}\ }\href {\doibase
  10.1103/PhysRevLett.113.231603} {\bibfield  {journal} {\bibinfo  {journal}
  {Phys. Rev. Lett.}\ }\textbf {\bibinfo {volume} {113}},\ \bibinfo {pages}
  {231603} (\bibinfo {year} {2014})},\ \Eprint {http://arxiv.org/abs/1409.3232}
  {arXiv:1409.3232 [hep-th]} \BibitemShut {NoStop}%
\bibitem [{\citenamefont {Xie}\ \emph {et~al.}(2021)\citenamefont {Xie},
  \citenamefont {Saffin},\ and\ \citenamefont {Zhou}}]{Xie:2021glp}%
  \BibitemOpen
  \bibfield  {author} {\bibinfo {author} {\bibfnamefont {Qi-Xin}\ \bibnamefont
  {Xie}}, \bibinfo {author} {\bibfnamefont {Paul~M.}\ \bibnamefont {Saffin}}, \
  and\ \bibinfo {author} {\bibfnamefont {Shuang-Yong}\ \bibnamefont {Zhou}},\
  }\bibfield  {title} {\enquote {\bibinfo {title} {{Charge-Swapping Q-balls and
  Their Lifetimes}},}\ }\href {\doibase 10.1007/JHEP07(2021)062} {\bibfield
  {journal} {\bibinfo  {journal} {JHEP}\ }\textbf {\bibinfo {volume} {07}},\
  \bibinfo {pages} {062} (\bibinfo {year} {2021})},\ \Eprint
  {http://arxiv.org/abs/2101.06988} {arXiv:2101.06988 [hep-th]} \BibitemShut
  {NoStop}%
\bibitem [{\citenamefont {Hou}\ \emph {et~al.}(2022)\citenamefont {Hou},
  \citenamefont {Saffin}, \citenamefont {Xie},\ and\ \citenamefont
  {Zhou}}]{Hou:2022jcd}%
  \BibitemOpen
  \bibfield  {author} {\bibinfo {author} {\bibfnamefont {Si-Yuan}\ \bibnamefont
  {Hou}}, \bibinfo {author} {\bibfnamefont {Paul~M.}\ \bibnamefont {Saffin}},
  \bibinfo {author} {\bibfnamefont {Qi-Xin}\ \bibnamefont {Xie}}, \ and\
  \bibinfo {author} {\bibfnamefont {Shuang-Yong}\ \bibnamefont {Zhou}},\
  }\bibfield  {title} {\enquote {\bibinfo {title} {{Charge-swapping Q-balls in
  a logarithmic potential and Affleck-Dine condensate fragmentation}},}\ }\href
  {\doibase 10.1007/JHEP07(2022)060} {\bibfield  {journal} {\bibinfo  {journal}
  {JHEP}\ }\textbf {\bibinfo {volume} {07}},\ \bibinfo {pages} {060} (\bibinfo
  {year} {2022})},\ \Eprint {http://arxiv.org/abs/2202.08392} {arXiv:2202.08392
  [hep-ph]} \BibitemShut {NoStop}%
\bibitem [{\citenamefont {Saffin}\ \emph {et~al.}(2022)\citenamefont {Saffin},
  \citenamefont {Xie},\ and\ \citenamefont {Zhou}}]{Saffin:2022tub}%
  \BibitemOpen
  \bibfield  {author} {\bibinfo {author} {\bibfnamefont {Paul~M.}\ \bibnamefont
  {Saffin}}, \bibinfo {author} {\bibfnamefont {Qi-Xin}\ \bibnamefont {Xie}}, \
  and\ \bibinfo {author} {\bibfnamefont {Shuang-Yong}\ \bibnamefont {Zhou}},\
  }\bibfield  {title} {\enquote {\bibinfo {title} {{Q-ball Superradiance}},}\
  }\href@noop {} {\  (\bibinfo {year} {2022})},\ \Eprint
  {http://arxiv.org/abs/2212.03269} {arXiv:2212.03269 [hep-th]} \BibitemShut
  {NoStop}%
\bibitem [{\citenamefont {Cardoso}\ \emph {et~al.}(2023)\citenamefont
  {Cardoso}, \citenamefont {Vicente},\ and\ \citenamefont
  {Zhong}}]{cardosoSub}%
  \BibitemOpen
  \bibfield  {author} {\bibinfo {author} {\bibfnamefont {Vitor}\ \bibnamefont
  {Cardoso}}, \bibinfo {author} {\bibfnamefont {Rodrigo}\ \bibnamefont
  {Vicente}}, \ and\ \bibinfo {author} {\bibfnamefont {Zhen}\ \bibnamefont
  {Zhong}},\ }\bibfield  {title} {\enquote {\bibinfo {title} {{On energy
  extraction from Q-balls and other fundamental solitons}},}\ }\href@noop {}
  {\bibfield  {journal} {\bibinfo  {journal} {unpublished}\ } (\bibinfo {year}
  {2023})}\BibitemShut {NoStop}%
\bibitem [{\citenamefont {Masson}\ and\ \citenamefont
  {Asenjo-Garcia}(2022)}]{DickeSuperNat}%
  \BibitemOpen
  \bibfield  {author} {\bibinfo {author} {\bibfnamefont {Stuart~J.}\
  \bibnamefont {Masson}}\ and\ \bibinfo {author} {\bibfnamefont {Ana}\
  \bibnamefont {Asenjo-Garcia}},\ }\bibfield  {title} {\enquote {\bibinfo
  {title} {{Universality of Dicke superradiance in arrays of quantum
  emitters}},}\ }\href@noop {} {\bibfield  {journal} {\bibinfo  {journal}
  {Nature Communications}\ }\textbf {\bibinfo {volume} {13}},\ \bibinfo {pages}
  {2285} (\bibinfo {year} {2022})}\BibitemShut {NoStop}%
\bibitem [{\citenamefont {Zhang}\ and\ \citenamefont {Zhou}()}]{zhangzhou}%
  \BibitemOpen
  \bibfield  {author} {\bibinfo {author} {\bibfnamefont {Guo-Dong}\
  \bibnamefont {Zhang}}\ and\ \bibinfo {author} {\bibfnamefont {Shuang-Yong}\
  \bibnamefont {Zhou}},\ }\bibfield  {title} {\enquote {\bibinfo {title} {{
  }},}\ }\href@noop {} {\bibinfo  {journal} {in preparation}\ }\BibitemShut
  {NoStop}%
\end{thebibliography}%

\end{document}